\documentclass[jgr]{agutex}

\usepackage{lineno}
\usepackage{rotating}
\usepackage[centertags]{amsmath}
\usepackage{bm}
\usepackage{url}

\usepackage{graphics}

\authorrunninghead{AAGAARD ET AL.}

\titlerunninghead{FAULT SLIP VIA DOMAIN DECOMPOSITION}

\authoraddr{B. T. Aagaard, USGS MS977, 345 Middlefield Rd, Menlo Park,
  CA 94025, USA. (baagaard@usgs.gov)}

\authoraddr{M. G. Knepley, Computation Institute, University of
  Chicago, Searle Chemistry Laboratory, 5735 S. Ellis Avenue, Chicago,
  IL 60637, USA.}

\authoraddr{C. A. Williams, GNS Science, 1 Fairway Drive, Avalon, PO
  Box 30368, Lower Hutt 5040, New Zealand.}

\begin{document}

\begin{quote}{\it
  This information is distributed solely for the purpose of
  predissemination peer review and must not be disclosed, released, or
  published until after approval by the U.S. Geological Survey
  (USGS). It is deliberative and predecisional information and the
  findings and conclusions in the document have not been formally
  approved for release by the USGS. It does not represent and should
  not be construed to represent any USGS determination or policy.
}\end{quote}

\title{A Domain Decomposition Approach to Implementing Fault Slip in
  Finite-Element Models of Quasi-static and Dynamic Crustal
  Deformation}

\authors{B. T. Aagaard,\altaffilmark{1}
  M. G. Knepley,\altaffilmark{2}
  and C. A. Williams\altaffilmark{3}}

\altaffiltext{1}{Earthquake Science Center, U.S. Geological Survey,
  Menlo Park, California, USA.}

\altaffiltext{2}{Computation Institute, University of Chicago,
  Chicago, Illinois, USA.}

\altaffiltext{3}{GNS Science, Lower Hutt, New Zealand.}

%
%
%

\begin{abstract}
  We employ a domain decomposition approach with Lagrange multipliers
  to implement fault slip in a finite-element code, PyLith, for use in
  both quasi-static and dynamic crustal deformation applications. This
  integrated approach to solving both quasi-static and dynamic
  simulations leverages common finite-element data structures and
  implementations of various boundary conditions, discretization
  schemes, and bulk and fault rheologies.  We have developed a custom
  preconditioner for the Lagrange multiplier portion of the system of
  equations that provides excellent scalability with problem size
  compared to conventional additive Schwarz methods. We demonstrate
  application of this approach using benchmarks for both quasi-static
  viscoelastic deformation and dynamic spontaneous rupture propagation
  that verify the numerical implementation in PyLith. 
\end{abstract}
  
\begin{article}

\section{Introduction}

The earthquake cycle, from slow deformation associated with
interseismic behavior to rapid deformation associated with earthquake
rupture, spans spatial scales ranging from fractions of a meter
associated with the size of contact asperities on faults and
individual grains to thousands of kilometers associated with plate
boundaries. Similarly, temporal scales range from fractions of a
second associated with slip at a point during earthquake rupture to
thousands of years of strain accumulation between earthquakes. The
complexity of the many physical processes operating over this vast
range of scales leads most researchers to focus on a narrow space-time
window to isolate just one or a few processes; the limited spatial and
temporal coverage of observations also often justifies this narrow
focus.

Researchers have recognized for some time, though, that interseismic
deformation and fault interactions influence earthquake rupture
propagation, and the dynamics of rupture propagation, in turn, affect
postseismic deformation \citep{Igarashi:etal:2003,Ito:etal:2007,Chen:Lapusta:2009,Matsuzawa:etal:2010}. In most cases one simplifies
some portion of the process to expedite the modeling results of
another portion. For example, studies of slow deformation associated
with interseismic and postseismic behavior often approximate dynamic
rupture behavior with the static coseismic slip
\citep{Reilinger:etal:2000,Pollitz:etal:2001,Langbein:etal:2006,Chlieh:etal:2007}.
Likewise, studies of rapid deformation associated with earthquake
rupture propagation often approximate the loading of the crust at the
beginning of a rupture
\citep{Mikumo:etal:1998,Harris:Day:1999,Aagaard:etal:BSSA:2001,Peyrat:etal:2001,Oglesby:Day:2001,Dunham:Archuleta:2004}. Numerical
seismicity models that attempt to model multiple earthquake cycles
generally simplify not only the fault loading and rupture propagation,
but also the physical properties to make the calculations tractable
\citep{Ward:1992,Robinson:Benites:1995,Hillers:etal:2006,Rundle:etal:2006,Pollitz:Schwartz:2008,Dieterich:Richards-Dinger:2010}.

Some dynamic spontaneous rupture modeling studies have
attempted to examine a broader space-time window to remove
simplifying assumptions and more accurately capture the complex
interactions over the earthquake cycle. For example,
\citet{Duan:Oglesby:2005} simulated multiple earthquake cycles on a
fault with a bend to capture the spatial variation in the
stress field around the bend, which they found to have a strong role
in determining whether a rupture would propagate past the bend. By
spinning up the model over many earthquake cycles, they obtained a
much more realistic stress field immediately prior to rupture compared
with assuming a simple stress field or calculating the stress field
from a static analysis. \citet{Chen:Lapusta:2009} examined the
behavior of small repeating earthquakes by modeling a stable sliding
region (friction increases with slip rate) surrounding an unstable
sliding region (friction decreases with slip rate). They found that
the aseismic slip occurring within the unstable patch between ruptures
contributed a significant fraction of the long-term slip. As a result,
their simulations displayed a complex interaction between aseismic
slip between earthquakes and coseismic slip that would not have been
possible if they did not explicitly model the interseismic
deformation. 

\citet{Kaneko:etal:2011} developed more sophisticated earthquake cycle
models using spectral element simulations that permit spatial
variations in physical properties that capture the dynamic rupture
propagation as well as the interseismic deformation. They examined the
effects of low-rigidity layers and a fault damaged zone on rupture
dynamics. In addition to purely dynamic effects, such as amplified
slip rates during dynamic rupture, they found several effects that
required resolving both the interseismic deformation and the rapid
slip during dynamic rupture; the low-rigidity layers reduced the
nucleation size, amplified slip rates during dynamic rupture,
increased the recurrence interval, and reduced the amount of aseismic
slip.

Reproducing observed earthquake cycle behavior remains a challenge.
\citet{Barbot:etal:2012} applied boundary integral simulation
techniques to develop an earthquake cycle model of Mw 6.0 Parkfield,
California, earthquakes. They employed spatial variation of the fault
constitutive properties for Dieterich-Ruina rate-state friction to
yield regions with stable sliding and regions with stick-slip
behavior. This allowed their numerical model to closely match the
observed geodetic interseismic behavior as well as the slip pattern of
the 2004 Parkfield earthquake. Nevertheless, some aspects of the
physical process, such as the 3-D nonplanar flower-structure geometry
of the San Andreas fault and 3-D variation in elastic properties were
not included in the \citet{Barbot:etal:2012} model.

Collectively, these studies suggest a set of desirable features for
models of the earthquake cycle to capture both the slow
deformation associated with interseismic behavior and the rapid
deformation associated with earthquake rupture propagation. These
features include the general capabilities of modeling elasticity with
elastic, viscoelastic, and viscoelastoplastic rheologies, as well as
slip on faults via either prescribed ruptures or spontaneous ruptures
controlled by a fault constitutive model. Additionally, a model could
also include the coupling of elasticity to fluid and/or heat flow.

With the goal of modeling the entire earthquake cycle with as few
simplifications as possible, much of our work in developing PyLith has
focused on modeling fault slip with application to quasi-static
simulations of interseismic and coseismic deformation and dynamic
simulations of earthquake rupture propagation. This effort builds on
our previous work on developing the numerical modeling software EqSim
\citep{Aagaard:etal:BSSA:2001} for dynamic spontaneous rupture
simulations and Tecton \citep{Melosh:Raefsky:1980,Williams:Richardson:1991} for
quasi-static interseismic and postseismic simulations. We plan to
seamlessly couple these two types of simulations together to resolve
the earthquake cycle. Implementing slip on the potentially nonplanar
fault surface differentiates these types of problems from many other
elasticity problems. Complexities arise because earthquakes may
involve offset on multiple, intersecting irregularly shaped fault
surfaces in the interior of a modeling domain. Furthermore, we want
the flexibility to either prescribe the slip on the fault or have the
fault slip evolve according to a fault constitutive model that
specifies the friction on the fault surface. Here, we describe a
robust, yet flexible method for implementing fault slip with a domain
decomposition approach, its effect on the overall design of PyLith,
and verification of its implementation using benchmarks.

\section{Numerical Model of Fault Slip}

In this section we summarize the formulation of the governing
equations using the finite-element method. We augment the conventional
finite-element formulation for elasticity with a domain decomposition
approach \citep{Smith:etal:1996,Zienkiewicz:etal:2005} to implement
the fault slip.  The PyLith manual \citep{PyLith:manual:1.7.1} provides
a step-by-step description of the formulation.

We solve the elasticity equation including inertial terms,
\begin{linenomath*}\begin{gather}
  \rho \frac{\partial^2\bm{u}}{\partial t^2} - \bm{f} 
  - \pmb{\nabla} \cdot \bm{\sigma} = \bm{0} \text{ in }V, \\
  \label{eqn:bc:Neumann}
  \bm{\sigma} \cdot \bm{n} = \bm{T} \text{ on }S_T, \\
  \label{eqn:bc:Dirichlet}
  \bm{u} = \bm{u}_0 \text{ on }S_u, \\
  \bm{d} - (\bm{u}_{+} - \bm{u}_{-}) = \bm{0}
  \text{ on }S_f, \label{eqn:fault:disp}
\end{gather}\end{linenomath*}
where $\bm{u}$ is the displacement vector, $\rho$ is the mass density,
$\bm{f}$ is the body force vector, $\bm{\sigma}$ is the Cauchy stress
tensor, and $t$ is time. We specify tractions $\bm{T}$ on surface
$S_T$, displacements $\bm{u_0}$ on surface $S_u$, and slip $\bm{d}$ on
fault surface $S_f$, where the tractions and fault slip are in global
coordinates. Because both $\bm{T}$ and $\bm{u}$ are vector quantities,
there can be some spatial overlap of the surfaces $S_T$ and $S_u$;
however, a degree of freedom at any location cannot be associated with
both prescribed displacements (Dirichlet) and traction (Neumann)
boundary conditions simultaneously.

Following a conventional finite-element formulation (ignoring the
fault surface for a moment), we construct the weak form by taking the
dot product of the governing equation with a weighting function and
setting the integral over the domain equal to zero,
\begin{linenomath*}\begin{equation}
  \int_{V} \pmb{\phi} \cdot 
  \left( \pmb{\nabla} \cdot \bm{\sigma} + \bm{f} -
    \rho\frac{\partial^{2}\bm{u}}{\partial t^{2}} \right) 
  \, dV=0.
\end{equation}\end{linenomath*}
The weighting function $\pmb{\phi}$ is a piecewise differentiable
vector field with $\pmb{\phi} = \bm{0}$ on $S_u$. After some algebra
and use of the boundary conditions (equations~(\ref{eqn:bc:Neumann})
and~(\ref{eqn:bc:Dirichlet})), we have
\begin{linenomath*}\begin{equation}
  \begin{split}
    - \int_{V} \nabla \pmb{\phi} : \bm{\sigma} \, dV
    + \int_{S_T} \pmb{\phi} \cdot \bm{T} \, dS
    + \int_{V} \pmb{\phi} \cdot \bm{f} \, dV \\
    - \int_{V} \pmb{\phi} \cdot \rho \frac{\partial^{2}\bm{u}}{\partial t^{2}} \, dV
    =0,
  \end{split}
\end{equation}\end{linenomath*}
where $\nabla \pmb{\phi} : \bm{\sigma}$ is the double inner product of
the gradient of the weighting function and the stress tensor.

\begin{figure}[h]
  \noindent\includegraphics{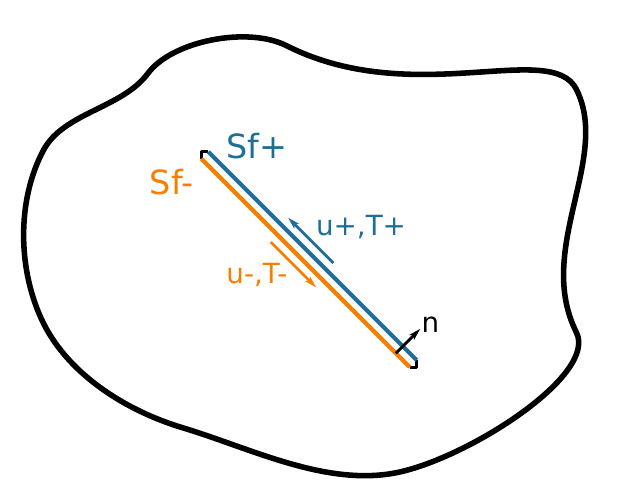}
  \caption{Diagram of domain decomposition approach for modeling fault
    slip. The fault slip introduces a jump in the displacement field
    across the fault, whereas the tractions are continuous.}
  \label{fig:domain:decomposition}
\end{figure}

Using a domain decomposition approach, we consider the fault surface
as an interior boundary between two domains as shown in
Figure~\ref{fig:domain:decomposition}. We assign a fault normal
direction to this interior boundary and ``positive'' and ``negative''
labels to the two sides of the fault, such that the fault normal is
the vector from the negative side of the fault to the positive side of
the fault. Slip on the fault is the displacement of the positive side
relative to the negative side. Slip on the fault also corresponds to
equal and opposite tractions on the positive ($\bm{l_{+}}$) and negative
($\bm{l_{-}}$) sides of the fault, which we impose using Lagrange
multipliers with $\bm{l}_{+} + \bm{l}_{-} = 0$.

Recognizing that the tractions on the fault surface are analogous to
the boundary tractions, we add in the contributions from integrating
the Lagrange multipliers (fault tractions) over the fault surface,
\begin{linenomath*}\begin{equation}
  \begin{split}
    - \int_{V} \nabla\pmb{\phi} : \bm{\sigma} \, dV
    + \int_{S_T} \pmb{\phi} \cdot \bm{T} \, dS
    - \int_{S_{f^+}} \pmb{\phi} \cdot \bm{l} \, dS \\
    + \int_{S_{f^-}} \pmb{\phi} \cdot \bm{l} \, dS
    + \int_{V} \pmb{\phi} \cdot \bm{f} \, dV 
    - \int_{V} \pmb{\phi} \cdot \rho \frac{\partial^{2}\bm{u}}{\partial t^{2}} \, dV
    =0.
  \end{split}
\end{equation}\end{linenomath*}
Our sign convention for the fault normal and fault tractions (tension
is positive) leads to the Lagrange multiplier terms having the
opposite sign as the boundary tractions term. We also construct the
weak form for the constraint associated with slip on the fault by
taking the dot product of the constraint equation with the weighting
function and setting the integral over the fault surface to zero,
\begin{linenomath*}\begin{equation}
  \int_{S_f} \pmb{\phi} \cdot 
  \left(\bm{d} - \bm{u}_{+} + \bm{u}_{-} \right) \, dS = 0.
\end{equation}\end{linenomath*}
This constraint equation applies to the relative displacement vector
across the fault and slip in the tangential and fault opening
directions.

The domain decomposition approach for imposing fault slip or tractions
on a fault is similar to the ``split nodes'' and ``traction at split
nodes'' (TSN) techniques used in a number of finite-difference and
finite-element codes
\citep{Melosh:Raefsky:1981,Andrews:1999,Bizzarri:Cocco:2005,Day:etal:2005,Duan:Oglesby:2005,Dalguer:Day:2007,Moczo:etal:2007},
but differs from imposing fault slip via double couple point
sources. The domain decomposition approach treats the fault surface as
a frictional contact interface, and the tractions correspond directly
to the Lagrange multipliers needed to satisfy the constraint equation
involving the jump in the displacement field across the fault and the
fault slip. As a result, the fault tractions are equal and opposite on
the two sides of the fault and satisfy equilibrium.  The TSN technique
is often applied in dynamic spontaneous rupture models with explicit
time stepping and a diagonal system Jacobian, so that the fault
tractions are explicitly computed as part of the solution of the
uncoupled equations. In this way the TSN technique as described by
\citet{Andrews:1999} could be considered an optimization of the domain
decomposition technique for the special case of dynamic spontaneous
rupture with a fault constitutive model and explicit time stepping.

Imposing fault slip via double couple point sources involves imposing
body forces consistent with an effective plastic strain associated
with fault slip (sometimes called the ``stress-free strain''
\citep{Aki:Richards:2002}). The total strain is the superposition of
this effective plastic strain and the elastic strain. The fault
tractions are associated with the elastic strain. This illustrates a
key difference between this approach and the domain decomposition
approach in which the Lagrange multipliers and the constraint equation
directly relate the fault slip to the fault tractions (Lagrange
multipliers). One implication of this difference is that when using
double couple point forces, the body forces driving slip depend on the
elastic modulii and will differ across a fault surface with a contrast
in the elastic modulii, whereas the fault tractions (Lagrange
multipliers) in the domain decomposition approach will be equal in
magnitude across the fault.

We express the weighting function $\pmb{\phi}$, trial solution
$\bm{u}$, Lagrange multipliers $\bm{l}$, and fault slip $\bm{d}$ as
linear combinations of basis functions,
\begin{linenomath*}\begin{gather}
\pmb{\phi} = \sum_{m} \bm{a}_m N_m, \\
\bm{u} = \sum_{n} \bm{u}_n N_n, \\
\bm{l} = \sum_{p} \bm{l}_p N_p, \\
\bm{d} = \sum_{p} \bm{d}_p N_p.
\end{gather}\end{linenomath*}
Because the weighting function is zero on $S_u$, the number of basis
functions for the trial solution $\bm{u}$ is generally greater than
the number of basis functions for the weighting function $\pmb{\phi}$,
i.e., $n > m$. The basis functions for the Lagrange multipliers and
fault slip are associated with the fault surface, which is a lower
dimension than the domain, so $p \ll n$ in most cases. If we express
the linear combination of basis functions in terms of a matrix-vector
product, we have
\begin{linenomath*}\begin{gather}
\pmb{\phi} = \bm{N}_m \cdot \bm{a}_m, \\
\bm{u} = \bm{N}_n \cdot \bm{u}_n, \\
\bm{l} = \bm{N}_p \cdot \bm{l}_p, \\
\bm{d} = \bm{N}_p \cdot \bm{d}_p.
\end{gather}\end{linenomath*}
The first term on the right hand side of these equations is a matrix
of the basis functions. For example, in three dimensions $\bm{N}_m$ is
a $3 \times 3m$ matrix, where $m$ is the number of basis functions.

The weighting function is arbitrary, so the integrands must be zero
for all $\bm{a}_m$, which leads to
\begin{linenomath*}\begin{gather}
  \label{eqn:residual:elasticity}
  \begin{split}
- \int_{V} \nabla \bm{N}_m^T \cdot \bm{\sigma} \, dV
+ \int_{S_T} \bm{N}_m^T \cdot \bm{T} \, dS
- \int_{S_{f^+}} \bm{N}_m^T \cdot \bm{N}_p \cdot \bm{l}_p \, dS \\
+ \int_{S_{f^-}} \bm{N}_m^T \cdot \bm{N}_p \cdot \bm{l}_p \, dS
+ \int_{V} \bm{N}_m^T \cdot \bm{f} \, dV \\
- \int_{V} \rho \bm{N}_m^T \cdot \bm{N}_n \cdot \frac{\partial^2 \bm{u}_n}{\partial
  t^2} \, dV
=\bm{0},
\end{split}
\\
  \label{eqn:residual:constraint}
  \int_{S_f} \bm{N}_p^T \cdot 
  \left( \bm{N}_p \cdot \bm{d}_p
    - \bm{N}_{n^+} \cdot \bm{u}_{n^+} 
    + \bm{N}_{n^-} \cdot \bm{u}_{n^-}
    \right) \, dS = \bm{0}.
\end{gather}\end{linenomath*}
We want to solve these equations for the coefficients $\bm{u}_n$
and $\bm{l}_p$ subject to $\bm{u} = \bm{u}_0 \text{ on
}S_u$. When we prescribe the slip, we specify $\bm{d}$ on $S_f$,
and when we use a fault constitutive model we specify how the
Lagrange multipliers $\bm{l}$ depend on the fault slip, slip rate,
and state variables.

We evaluate the integrals in equations~(\ref{eqn:residual:elasticity})
and~(\ref{eqn:residual:constraint}) using numerical quadrature
\citep{Zienkiewicz:etal:2005}. This involves evaluating the
integrands at the quadrature points, multiplying by the corresponding
weighting function, and summing over the quadrature points. With an
appropriate choice for the quadrature scheme the finite-element method
allows inclusion of spatial variations of boundary tractions, density,
body forces, and physical properties within the cells.

To solve equations~(\ref{eqn:residual:elasticity})
and~(\ref{eqn:residual:constraint}), we construct a linear system of
equations.  For nonlinear bulk rheologies it is convenient to work
with the increment in stress and strain, so we formulate the solution
of the equations in terms of the increment in the solution from time
$t$ to $t+\Delta t$ rather than the solution at time $t+\Delta t$.
Consequently, rather than constructing a system with the form $\bm{A}
\cdot \bm{u}(t+\Delta t) = \bm{b}(t+\Delta t)$, we construct a system
with the form $\bm{A} \cdot \bm{du} = \bm{b}(t+\Delta t) - \bm{A}
\cdot \bm{u}(t)$, where $\bm{u}(t+\Delta t) = \bm{u}(t) + \bm{du}$. We
use an initial guess of zero for the increment in the solution.

\subsection{Quasi-static Simulations}

For quasi-static simulations we ignore the inertial term and
time-dependence only enters through the constitutive models and the
loading conditions. As a result, the quasi-static simulations are a
series of static problems with potentially time-varying physical
properties and boundary conditions. The temporal accuracy of the
solution is limited to resolving these temporal
variations. Considering the deformation at time $t+\Delta t$,
\begin{linenomath*}\begin{gather}
  \label{eqn:quasi-static:residual:elasticity}
  \begin{split}
    - \int_{V} \nabla \bm{N}_m^T \cdot \bm{\sigma}(t+\Delta t) \, dV
    + \int_{S_T} \bm{N}_m^T \cdot \bm{T}(t+\Delta t) \, dS \\
    - \int_{S_{f^+}} \bm{N}_m^T \cdot \bm{N}_p \cdot \bm{l}_p(t+\Delta t) \, dS \\
    + \int_{S_{f^-}} \bm{N}_m^T \cdot \bm{N}_p \cdot \bm{l}_p(t+\Delta t) \, dS
    \\
    + \int_{V} \bm{N}_m^T \cdot \bm{f}(t+\Delta t) \, dV
    =\bm{0},
  \end{split}
  \\
\label{eqn:quasi-static:residual:fault}
  \begin{split}
    \int_{S_f} \bm{N}_p^T \cdot
    \left( \bm{N}_p \cdot \bm{d}_p(t+\Delta t)
      - \bm{N}_{n^+} \cdot \bm{u}_{n^+}(t+\Delta t) 
    \right) \, dS \\ +\int_{S_f} \bm{N}_p^T \cdot \left(
      \bm{N}_{n^-} \cdot \bm{u}_{n^-}(t+\Delta t)
    \right) \, dS = \bm{0}.
  \end{split}
\end{gather}\end{linenomath*}
To march forward in time, we simply increment time, solve the
equations, and add the increment in the solution to the solution from
the previous time step.  We solve these equations using the Portable,
Extensible Toolkit for Scientific Computation (PETSc), which provides
a suite of tools for solving linear systems of algebraic equations
with parallel processing
\citep{PETSC:efficient,PETSc:manual}. In solving the
system, we compute the residual (i.e., $\bm{r} = \bm{b} -
\bm{A} \cdot \bm{u}$) and the Jacobian of the system
($\bm{A}$). In our case the solution is $\bm{u} =
\left( \begin{smallmatrix} \bm{u}_n \\
    \bm{l}_n \end{smallmatrix} \right)$, and the residual is
simply the left sides of
equations~(\ref{eqn:quasi-static:residual:elasticity})
and~(\ref{eqn:quasi-static:residual:fault}). 

The Jacobian of the system, $\bm{A}$, is the action (operation) that
we apply to the increment of the solution, $\bm{du}$.  To find the
portion of the Jacobian associated with
equation~(\ref{eqn:quasi-static:residual:elasticity}), we let
$\bm{\sigma}(t+\Delta t) = \bm{\sigma}(t) + \bm{d\sigma}(t)$. The
action on the increment of the solution is associated with the
increment in the stress tensor $\bm{d\sigma}(t)$. We approximate the
increment in the stress tensor using linear elasticity and
infinitesimal strains,
\begin{linenomath*}\begin{equation}
  \bm{d\sigma}(t) = \frac{1}{2} \bm{C}(t) \cdot (\nabla + \nabla^T)
  \bm{u}(t),
\end{equation}\end{linenomath*}
where $\bm{C}$ is the fourth order tensor of elastic constants. For
bulk constitutive models with a linear response, $\bm{C}$ is constant
in time. For other constitutive models we form $\bm{C}(t)$ from the
current solution and state variables. Substituting into the first term
in equation~(\ref{eqn:quasi-static:residual:elasticity}) and
expressing the displacement vector as a linear combination of basis
functions, we find this portion of the Jacobian is
\begin{linenomath*}\begin{equation}\label{eqn:jacobian:implicit:stiffness}
  \bm{K} = \frac{1}{4} \int_V 
  (\nabla^T + \nabla) \bm{N}_m^T \cdot
  \bm{C} \cdot (\nabla + \nabla^T) \bm{N}_n  \, dV.
\end{equation}\end{linenomath*}
This matches the tangent stiffness matrix in conventional solid
mechanics finite-element formulations. In computing the residual, we
use the expression given in equation~(\ref{eqn:residual:elasticity})
with one implementation for infinitesimal strain and another
implementation for small strain and rigid body motion using the
Green-Lagrange strain tensor and the second Piola-Kirchoff stress
tensor \citep{Bathe:1995}. Following a similar procedure, we find the
portion of the Jacobian associated with the constraints,
equation~(\ref{eqn:quasi-static:residual:fault}), is
\begin{linenomath*}\begin{equation}\label{eqn:jacobian:constraint}
  \bm{L} = \int_{S_f} \bm{N}_p^T \cdot (\bm{N}_{n^+} - \bm{N}_{n^-}) \, dS.
\end{equation}\end{linenomath*}
Thus, the Jacobian of the entire system has the form,
\begin{linenomath*}\begin{equation}\label{eqn:saddle:point}
  \bm{A} = 
  \left( \begin{array}{cc}
      \bm{K} & \bm{L}^T \\ \bm{L} & \bm{0} 
    \end{array} \right).
\end{equation}\end{linenomath*}
Note that the terms in $\bm{N_{n^+}}$ and $\bm{N_{n^-}}$ are
identical, but they refer to degrees of freedom (DOF) on the positive and
negative sides of the fault, respectively. Consequently, in practice
we compute the terms for the positive side of the fault and assemble
the terms into the appropriate DOF for both sides of
the fault. Hence, we compute
\begin{linenomath*}\begin{equation}\label{eqn:jacobian:constraint:code}
  \bm{L_p} = \int_{S_f} \bm{N}_p^T \cdot \bm{N}_{n^+} \, dS,
\end{equation}\end{linenomath*}
with the Jacobian of the entire system taking the form,
\begin{linenomath*}\begin{equation}\label{eqn:saddle:point:code}
  \bm{A} = 
  \left( \begin{array}{cccc}
      \bm{K}_{nn} & \bm{K}_{nn^+} & \bm{K}_{nn^-} & \bm{0} \\
      \bm{K}_{n^+n} & \bm{K}_{n^+n^+} & \bm{0} & \bm{L}_p^T \\ 
      \bm{K}_{n^-n} & \bm{0} & \bm{K}_{n^-n^-} & -\bm{L}_p^T \\ 
      \bm{0} & \bm{L}_p & -\bm{L}_p & \bm{0} 
    \end{array} \right),
\end{equation}\end{linenomath*}
where $n$ denotes DOF not associated with the fault,
$n^-$ denotes DOF associated with the negative side of
the fault, $n^+$ denotes DOF associated with the
positive side of the fault, and $p$ denotes DOF
associated with the Lagrange multipliers.

The matrix $\bm{L}$ defined in
equation~(\ref{eqn:jacobian:constraint}) is spectrally equivalent to
the identity, because it involves integration of products of the basis
functions. This makes the traditional Ladyzhenskaya-Babuska-Brezzi
(LBB) stability criterion \citep{Brenner:Scott:2008} trivial to satisfy by
choosing the space of Lagrange multipliers to be exactly the space of
displacements, restricted to the fault. This means we simply need to
know the distance between any pair of vertices spanning the fault,
which can be expressed as a relative displacement, i.e., fault slip.

\subsection{Dynamic Simulations}

In dynamic simulations we include the inertial term to
resolve the propagation of seismic waves, with an intended focus on
applications for earthquake physics and ground-motion simulations. The
general form of the system Jacobian remains the same as in
quasi-static simulations given in
equation~(\ref{eqn:saddle:point}). The integral equation for the fault
slip constraint remains unchanged, so the corresponding portions of
the Jacobian ($\bm{L}$) and residual ($\bm{r}_p$) are also exactly the
same as in the quasi-static simulations. Including the inertial term
in equation~(\ref{eqn:quasi-static:residual:elasticity}) for time $t$
rather than $t+\Delta t$ yields
\begin{linenomath*}\begin{equation}\label{eqn:dynamic:residual:elasticity}
  \begin{split}
    - \int_{V} \nabla \bm{N}_m^T \cdot \bm{\sigma}(t) \, dV
    + \int_{S_T} \bm{N}_m^T \cdot \bm{T}(t) \, dS \\
    - \int_{S_{f^+}} \bm{N}_m^T \cdot \bm{N}_p \cdot \bm{l}_p(t) \, dS \\
    + \int_{S_{f^-}} \bm{N}_m^T \cdot \bm{N}_p \cdot \bm{l}_p(t) \, dS
    \\
    + \int_{V} \bm{N}_m^T \cdot \bm{f}(t) \, dV \\
      - \int_{V} \rho \bm{N}_m^T \cdot \bm{N}_n \cdot 
          \frac{\partial^2 \bm{u}_n(t)}{\partial t^2} \, dV
    =\bm{0}.
  \end{split}
\end{equation}\end{linenomath*}

We find the upper portion of the Jacobian of the system by considering
the action on the increment in the solution, just as we did for the
quasi-static simulations. In this case we associate the increment in
the solution with the temporal discretization. We march forward in
time using explicit time stepping via Newmark's method
\citep{Newmark:1959} with a central difference scheme wherein the
acceleration and velocity are given by
\begin{linenomath*}\begin{gather}
  \frac{\partial^2 \bm{u}(t)}{\partial t^2} = 
  \frac{1}{\Delta t^2} \left(
    \bm{du} - \bm{u}(t) + \bm{u}(t-\Delta t)
  \right), \\
  \frac{\partial \bm{u}(t)}{\partial t} = \frac{1}{2\Delta t} \left(
    \bm{du} + \bm{u}(t) - \bm{u}(t-\Delta t)
    \right).
\end{gather}\end{linenomath*}
Expanding the inertial term yields
\begin{linenomath*}\begin{equation}
  \begin{split}
    - \int_{V} \rho \bm{N}_m^T \cdot \bm{N}_n \cdot \frac{\partial^2 \bm{u}_n(t)}{\partial
      t^2} \, dV = \\
    - \frac{1}{\Delta t^2} \int_{V} \rho \bm{N}_m^T \cdot
    \bm{N}_n \cdot 
    \left( \bm{du}_n(t) - \bm{u}_n(t) + \bm{u}_n(t-\Delta t) \right) \, dV,
  \end{split}
\end{equation}\end{linenomath*}
so that the upper portion of the Jacobian is
\begin{linenomath*}\begin{equation}
  \label{eqn:jacobian:explicit:inertia}
  \bm{K} = 
    \frac{1}{\Delta t^2} \int_{V} \rho \bm{N}_m^T\ \cdot \bm{N}_n \, dV.
\end{equation}\end{linenomath*}
This matches the mass matrix in conventional solid mechanics
finite-element formulations.

Earthquake ruptures in which the slip has a short rise time tend to
introduce deformation at short length scales (high frequencies) that
numerical models cannot resolve accurately. This is especially true
in spontaneous rupture simulations, because the rise time is sensitive
to the evolution of the fault rupture. To reduce the
introduction of deformation at such short length scales we add
artificial damping via Kelvin-Voigt viscosity
\citep{Day:etal:2005,Kaneko:etal:2008} to the computation of the strain,
\begin{linenomath*}\begin{gather}
  \bm{\varepsilon} = \frac{1}{2} \left[ \nabla \bm{u} +
    (\nabla \bm{u})^T \right ], \\
  \bm{\varepsilon} \approx \frac{1}{2} \left[ \nabla \bm{u}_d +
    (\nabla \bm{u}_d)^T \right ], \\
  \bm{u_d} = \bm{u} + \eta^* \Delta t \frac{\partial
    \bm{u}}{\partial t},
\end{gather}\end{linenomath*}
where $\eta^*$ is a nondimensional viscosity on the order of
0.1--1.0. 

\subsection{Nondimensionalization}

The domain decomposition approach for implementing fault slip with
Lagrange multipliers requires solving for both the displacement field
and the Lagrange multipliers, which correspond to fault tractions. We
expect the displacements to be generally on the order of mm to m
whereas the fault tractions will be on the order of MPa. Thus,
if we use dimensioned quantities in SI units, then we would expect the
solution to include terms that differ by up to nine orders of
magnitude. This results in a rather ill-conditioned system. We avoid
this ill-conditioning by nondimensionalizing all of the quantities
involved in the problem based upon user-specified scales
\citep{PyLith:manual:1.7.1}, facilitating the formation of
well-conditioned systems of equations for problems across a wide range
of spatial and temporal scales.

\subsection{Prescribed Fault Rupture}

In a prescribed (kinematic) fault rupture we specify the slip-time
history $\bm{d}(x,y,z,t)$ at every location on the fault surfaces.
The slip-time history enters into the calculation of the residual as
do the Lagrange multipliers, which are available from the current
trial solution. In prescribing the slip-time history we do not specify
the initial tractions on the fault surface so the Lagrange multipliers
are the \textit{change} in the tractions on the fault surfaces
corresponding to the slip. PyLith includes a variety of slip-time
histories, including a step function, a linear ramp (constant slip
rate), the integral of Brune's far-field time function
\citep{Brune:1970}, a sine-cosine function developed by
\citet{Liu:etal:2006}, and a user-defined time history. These are
discussed in detail in the PyLith manual
\citep{PyLith:manual:1.7.1}. PyLith allows specification of the slip
initiation time independently at each location as well as
superposition of multiple earthquake ruptures with different origin
times, thereby permitting complex slip behavior.

\subsection{Spontaneous Fault Rupture}

In contrast to prescribed fault rupture, in spontaneous fault rupture
a constitutive model controls the tractions on the fault surface. The
fault slip evolves based on the fault tractions as driven by the
initial conditions, boundary conditions and deformation. In our
formulation of fault slip, slip is assumed to be known and the
fault tractions (Lagrange multipliers) are part of the solution
(unknowns). The fault constitutive model places bounds on the Lagrange
multipliers and the system of equations is nonlinear-- when a location
on the fault is slipping, we must solve for both the fault slip (which
is known in the prescribed ruptures) and the Lagrange multipliers to
find values consistent with the fault constitutive model.

At each time step, we first assume the increment in the fault slip is
zero, so that the Lagrange multipliers correspond to the fault
tractions required to lock the fault. If the Lagrange multipliers
exceed the fault tractions allowed by the fault constitutive model,
then we iterate to find the increment in slip that yields Lagrange
multipliers that satisfy the fault constitutive model.  On the other
hand, if the Lagrange multipliers do not exceed the fault tractions
allowed by the fault constitutive model, then the increment in fault
slip remains zero, and no adjustments to the solution are necessary.

In iterating to find the fault slip and Lagrange multipliers that
satisfy the fault constitutive model, we employ the following
procedure.  We use this same procedure for all fault constitutive
models, but it could be specialized to provide better performance
depending on how the fault constitutive model depends on slip, slip
rate, and various state variables. We first compute the perturbation
in the Lagrange multipliers necessary to satisfy the fault
constitutive model for the current estimate of slip. We then compute
the increment in fault slip corresponding to this perturbation in the
Lagrange multipliers assuming deformation is limited to vertices on
the fault. That is, we consider only the DOF associated with the fault
interface when computing how a perturbation in the Lagrange
multipliers corresponds to a change in fault slip. In terms of the
general form of a linear system of equations ($\bm{A} \bm{u} =
\bm{b}$), our subset of equations based on
equation~(\ref{eqn:saddle:point:code}) has the form
\begin{linenomath*}\begin{equation}
  \begin{pmatrix}
    \bm{K}_{n^+n^+} & 0 & \bm{L}_p^T  \\
    0 & \bm{K}_{n^-n^-} & -\bm{L}_p^T \\
    \bm{L}_p & -\bm{L}_p & 0
  \end{pmatrix}
  \begin{pmatrix}
  \bm{u}_{n^+} \\
  \bm{u}_{n^-} \\
  \bm{l}_p \\
  \end{pmatrix}
  =
  \begin{pmatrix}
  \bm{b}_{n^+} \\
  \bm{b}_{n^-} \\
  \bm{b}_p \\
  \end{pmatrix},
\end{equation}\end{linenomath*}
where $n^+$ and $n^-$ refer to the DOF associated with
the positive and negative sides of the fault,
respectively. Furthermore, we can ignore the terms $\bm{b}_{n^+}$
and $\bm{b}_{n^-}$ because they remain constant as we change the
Lagrange multipliers or fault slip. Our task reduces to solving the
following system of equations to estimate the change in fault slip
$\partial \bm{d}$ associated with a perturbation in the Lagrange
multipliers $\partial \bm{l}_p$:
\begin{linenomath*}\begin{gather}
  \label{eqn:spontaneous:rupture:update:lagrange}
  \bm{K}_{n^+n^+} \cdot \partial \bm{u}_{n^+} = 
  - \bm{L}_p^T \cdot \partial \bm{l}_p, \\
  \bm{K}_{n^-n^-} \cdot \partial \bm{u}_{n^-} =
  \bm{L}_p^T \cdot \partial \bm{l}_p, \\
  \label{eqn:spontaneous:rupture:update:slip}
  \partial \bm{d}_p =  \partial \bm{u}_{n^+} - \partial \bm{u}_{n^-}.
\end{gather}\end{linenomath*}

The efficiency of this iterative procedure depends on both the fault
constitutive model and how confined the deformation is to the region
immediately surrounding the fault. If the fault friction varies
significantly with slip, then the estimate of how much slip is
required to match the fault constitutive model will be
poor. Similarly, in rare cases in which the fault slip extends across
the entire domain, deformation extends far from the fault and
the estimate derived using only the fault DOF will be
poor. To make this iterative procedure more robust so that it
works well across a wide variety of fault constitutive models, we add
a small enhancement to the iterative procedure.

At each iteration we use a simple line search to find the
increment in slip that best satisfies the fault constitutive
model. Specifically, we search for $\alpha$ using a bilinear search in
logarithmic space to minimize
\begin{linenomath*}\begin{equation}
  C = \| \bm{l}_p + \alpha \partial\bm{l}_p - f(\bm{d}_p +
  \alpha \partial\bm{d}_p) \|_2,
\end{equation}\end{linenomath*}
where $f(\bm{d})$ corresponds to the fault constitutive model and
$\|x\|_2$ denotes the L$^2$-norm of $x$. Performing this search in
logarithmic space rather than linear space greatly accelerates the
convergence in constitutive models in which the coefficient of
friction depends on the logarithm of the slip rate.

PyLith includes several commonly used fault constitutive models, all
of which specify the shear traction on the fault $T_f$ as a function
of the cohesive stress $T_c$, coefficient of friction, $\mu_f$, and
normal traction $T_n$,
\begin{linenomath*}\begin{equation}
  T_f = T_c - \mu_f T_n.
\end{equation}\end{linenomath*}
$T_f$ in this equation corresponds to the magnitude of the shear
traction vector; the shear traction vector is resolved into the
direction of the slip rate. We use the sign convention that
compressive normal tractions are negative. When the fault is under
compression, we prevent interpenetration, and when the fault is under
tension, the fault opens ($d_n > 0$) and the fault traction vector is
zero. The fault constitutive models include static friction, linear
slip-weakening \citep{Ida:1972}, linear time-weakening
\citep{Andrews:2004}, and Dieterich-Ruina rate-state friction with an
aging law \citep{Dieterich:1979}. See the PyLith manual
\citep{PyLith:manual:1.7.1} for details.

\section{Finite-Element Mesh Processing}

Like all finite-element engines, PyLith performs operations on cells
and vertices comprising the discretized domain (finite-element
mesh). These operations include calculating cell and face integrals to
evaluate weak forms, assemble local cell vectors and matrices into
global vector and matrix objects, impose Dirichlet boundary conditions
on the algebraic system, and solve the resulting system of nonlinear
algebraic equations. In PyLith, these operations are accomplished
using PETSc, and in particular the Sieve package for finite-element
support \citep{Knepley:Karpeev:2009}.

\begin{figure*}[h]
  \noindent\includegraphics{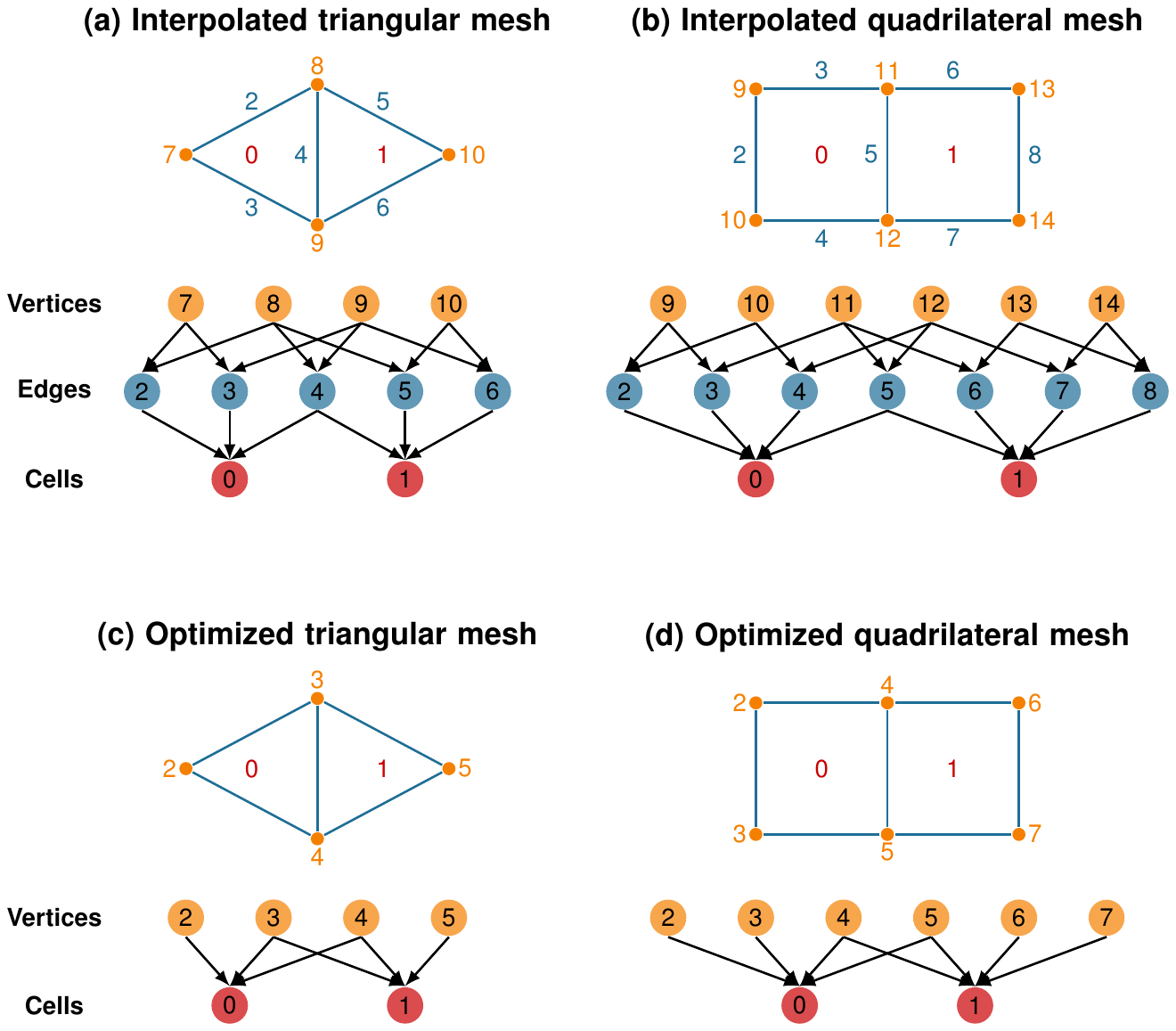}
  \caption{Direct acyclic graph representations of the covering
    relation for 2-D meshes with triangular and quadrilateral cells. The graphs for
    interpolated meshes (a) and (b) include all levels of the topology
    whereas the graphs for optimized meshes (c) and (d) only include
    the top and bottom levels. The graphs for interpolated meshes in
    3-D include faces between edges and cells. PyLith currently uses the
    optimized graph representations.}
  \label{fig:sieve}
\end{figure*}

The Sieve application programming interface (API) for mesh
representation and manipulation is based upon a direct acyclic graph
representation of the \textit{covering} relation in a mesh,
illustrated in Figure~\ref{fig:sieve}. For example, faces cover cells,
edges cover faces, and points cover edges. By focusing on the key
topological relations, the interface can be both concise and quite
general. Using this generic API, PyLith is able to support one, two,
and three dimensional meshes, with simplicial, hex, and even prismatic
cell shapes, while using very little dimension or shape specific
code. However, in order to include faults, we include additional
operations in Sieve beyond those necessary for conventional
finite-element operations.

In our domain decomposition approach, the finite-element mesh includes
the fault as an interior surface. This forces alignment of the element
faces along the fault. To impose a given fault slip as in
equation~(\ref{eqn:fault:disp}), we must represent the displacement on
both sides of the fault for any vertex on the fault. One option is to
designate ``fault vertices'' which possess twice as many displacement
DOF \citep{Aagaard:etal:BSSA:2001}. However, this
requires storing the global variable indices by cell rather than by
vertex or adding special fault metadata to the vertices, significantly
increasing storage costs and/or index lookup costs. 

We choose another option and modify the initial finite-element mesh by
replacing each fault face with a zero-volume cohesive cell.  Many mesh
generation tools do not support specification of faces on interior
surfaces. Consequently, we create these cohesive cells in a
preprocessing step at the beginning of a simulation. We construct the
set of oriented fault faces from a set of vertices marked as lying on
the fault. We join these vertices into faces, consistently orient them
(using a common fault normal direction), and associate them with pairs
of cells in the original mesh.

\begin{figure*}[h]
  \noindent\includegraphics{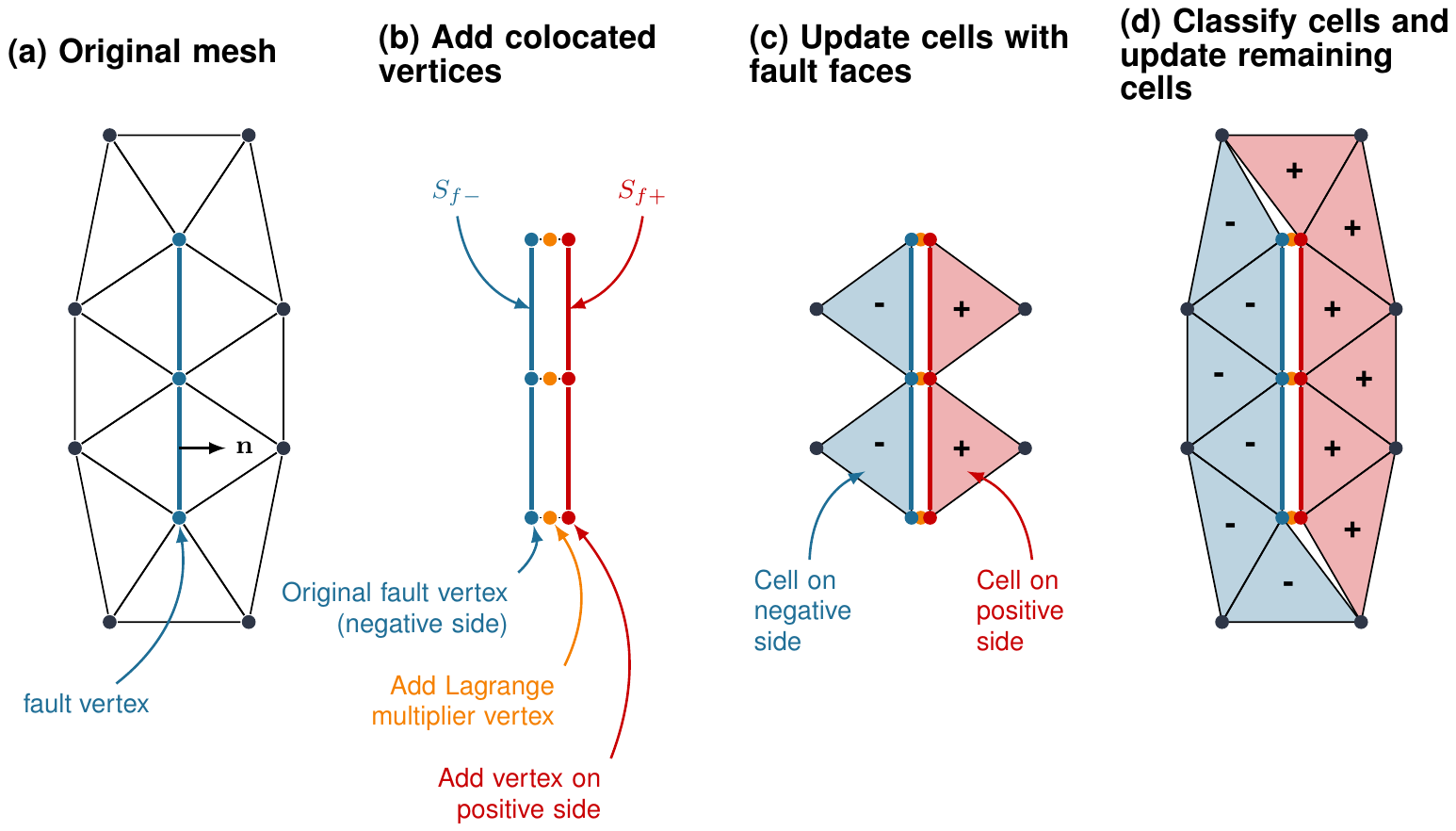}
  \caption{Construction of cohesive cells for a fault. (a) Original
    mesh with fault normal and fault vertices identified. (b) For each
    vertex on the fault, introduce a vertex on the positive side of
    the fault $S_{f^+}$ and a vertex corresponding to the Lagrange
    multiplier constraint between the pair of vertices on the positive
    and negative sides of the fault. (c) Identify cells with faces on
    the fault. Use the orientation of each face to identify cells on
    the positive and negative sides of the fault. Replace vertices in
    cells on the positive side of the fault with the newly created
    vertices. (d) Classify remaining cells with vertices on the fault
    using breadth-first search, and replace original vertices in cells
    on positive side of the fault with newly created
    vertices. Construct cohesive cells with zero volume from the
    vertices on the positive side of the fault, negative side of the
    fault, and Lagrange multiplier constraints.}
  \label{fig:cohesive:cell}
\end{figure*}

Given this set of oriented fault faces, we introduce a set of cohesive
cells using a step-by-step modification of the Sieve data structure
representing the mesh illustrated in
Figure~\ref{fig:cohesive:cell}. First, for each vertex on the negative
side of the fault $S_{f^-}$, we introduce a second vertex on the
positive side of the fault $S_{f^+}$ and a third vertex corresponding
to the Lagrange multiplier constraint. The Lagrange multiplier vertex
lies on an edge between the vertex on $S_{f^+}$ and the vertex on
$S_{f^-}$. The fault faces are organized as a Sieve, and each face has
the two cells it is associated with as descendants. Because the cells
are consistently oriented, the first cell attached to each face is on
the negative side of the fault, i.e., $S_{f^-}$. We
replace the vertices on the fault face of each second cell, which is
on the positive side of the fault, i.e., $S_{f^+}$, with the newly
created vertices. Finally, we add a cohesive cell including the
original fault face, a face with the newly created vertices, and the
Lagrange vertices. These cohesive cells are prisms. For example, in a
tetrahedral mesh the cohesive cells are triangular prisms, whereas in
a hexahedral meshes they are hexahedrons.

We must also update all cells on the positive side of the fault that
touch the fault with only an edge or single vertex. We need to replace
the original vertices with the newly introduced vertices on the
positive side of the fault. In cases where the fault reaches the
boundaries of the domain, it is relatively easy to identify these
cells because these vertices are shared with the cells that have faces
on the positive side of the fault. However, in the case of a fault
that does not reach the boundary of the domain, cells near the ends of
the fault share vertices with cells that have a face on the positive
side of the fault \textit{and} cells that have a face on the negative
side of the fault. We use a breadth-first classification scheme to
classify all cells with vertices on the fault into those having vertices
on the positive side of the fault and those having vertices on the
negative side of the fault, so that we can replace the original
vertices with the newly introduced vertices on the positive side of
the fault.

In classifying the cells we iterate over the set of fault
vertices. For each vertex we examine the set of cells attached to that
vertex, called the \emph{support} of the vertex in the Sieve API
\citep{Knepley:Karpeev:2009}. For each unclassified cell in the
support, we look at all of its neighbors that touch the fault. If any
is classified, we give the cell this same classification. If not, we
continue with a breadth-first search of its neighbors until a
classified cell is found. This search must terminate because there are
a finite number of cells surrounding the vertex and at least one is
classified (contains a face on the fault with this vertex). Depending
on the order of the iteration, this can produce a ``wrap around''
effect at the ends of the fault, but it does not affect the numerical
solution as long as the fault slip is forced to be zero at the edges of
the fault. In prescribed slip simulations this is done via the
user-specified slip distribution, whereas in spontaneous rupture
simulations it is done by preventing slip with artificially large
coefficients of friction, cohesive stress, or compressive normal tractions.

%

\section{Solver Customization}

\subsection{Quasi-static Simulations}
\label{sec:solver:quasi-static}

To solve the large, sparse systems of linear equations
arising in our quasi-static simulations, we employ preconditioned
Krylov subspace methods~\citep{Saad03}. We create a sequence of
vectors by repeatedly applying the system matrix to the
right-hand-side vector, $\bm{A^k} \cdot \bm{b}$, and they form
a basis for a subspace, termed the Krylov space. We can efficiently
find an approximate solution in this subspace.  Because sparse
matrix-vector multiplication is scalable via parallel processing, this
is the method of choice for parallel simulation. However, for most
physically relevant problems, the Krylov solver requires a
preconditioner to accelerate convergence. While generic
preconditioners exist~\citep{Saad03,Smith:etal:1996}, the method must
often be specialized to a particular problem. In this section we
describe a preconditioner specialized to our formulation for fault
slip with Lagrange multipliers.

The introduction of Lagrange multipliers to implement the fault slip
constraints produces the saddle-point problem shown in
equation~(\ref{eqn:saddle:point}). Traditional black-box parallel
preconditioners, such as the additive Schwarz Method (ASM)
\citep{Smith:etal:1996}, are not very effective for this type of
problem and produce slow convergence. However, PETSc provides tools to
construct many variations of effective parallel preconditioners for
saddle point problems.

The field split preconditioner in PETSc \citep{PETSc:manual} allows
the user to define sets of unknowns which correspond to different
fields in the physical problem. This scheme is flexible enough to
accommodate an arbitrary number of fields, mixed discretizations,
fields defined over a subset of the mesh, etc. Once these fields are
defined, a substantial range of preconditioners can be assembled using
only PyLith options for PETSc. Table~\ref{tab:preconditioner:options} shows
example preconditioners and the options necessary to construct them.

Another option involves using the field split preconditioner in PETSc
in combination with a custom preconditioner for the submatrix
associated with the Lagrange multipliers. In formulating the custom
preconditioner, we exploit the structure of the sparse Jacobian
matrix. Our system Jacobian has the form
\begin{linenomath*}\begin{equation}
  \bm{A} = \left( \begin{array}{cc}
      \bm{K} & \bm{L}^T \\
      \bm{L} & \bm{0}
    \end{array} \right).
\end{equation}\end{linenomath*}
The Schur complement $\bm{S}$ of the submatrix $\bm{K}$ is given by,
\begin{linenomath*}\begin{equation}
  \bm{S} = -\bm{L} \bm{K}^{-1} \bm{L}^T
\end{equation}\end{linenomath*}
which leads to a simple block diagonal preconditioner for $\bm{A}$
\begin{linenomath*}\begin{equation}
  \bm{P} = \left( \begin{array}{cc}
    \bm{P}_\mathit{elasticity} & 0 \\
    0 & \bm{P}_\mathit{fault}
  \end{array} \right)
  = \left( \begin{array}{cc}
    \bm{K} & 0 \\
    0 & -\bm{L} \bm{K}^{-1} \bm{L}^T
  \end{array} \right).
\end{equation}\end{linenomath*}

The elastic submatrix $\bm{K}$, in the absence of boundary conditions,
has three translational and three rotational null modes. These are
provided to the algebraic multigrid (AMG) preconditioner, such as the
ML library \citep{ML:users:guide} or the PETSc GAMG preconditioner, to
assure an accurate coarse grid solution. AMG mimics the action of
traditional geometric multigrid, but it generates coarse level
operators and interpolation matrices using only the system matrix,
treated as a weighted graph, rather than a separate description of the
problem geometry, such as a mesh. We split the elastic block from the
fault block and also manage the Schur complements. In this way, all
block preconditioners, including those nested with multigrid, can be
controlled from the options file without recompilation or special
code.

We now turn our attention to evaluating the fault portion of the
preconditioning matrix associated with the Lagrange multipliers, since
PETSc preconditioners can handle the elastic portion as
discussed in the previous paragraph. In computing
$\bm{P_\mathit{fault}}$ we approximate $\bm{K}^{-1}$ with
the inverse of the diagonal portion of $\bm{K}$. Because $\bm{L}$ 
consists of integrating the products of basis functions over the fault
faces, its structure depends on the quadrature scheme and the choice
of basis functions. For conventional low order finite-elements and
Gauss quadrature, $\bm{L}$ contains nonzero terms coupling the
degree of freedom for each coordinate axes of a vertex with the
corresponding degree of freedom of the other vertices in a
cell. However, if we collocate quadrature points at the cell vertices,
then only one basis function is nonzero at each quadrature point and
$\bm{L}$ becomes block diagonal; this is also true for spectral
elements with Legendre polynomials and Gauss-Lobatto-Legendre
quadrature points. This leads to a diagonal matrix for the lower
portion of the conditioning matrix,
\begin{linenomath*}\begin{equation}
  \bm{P}_\mathit{fault} = -\bm{L}_p (\bm{K}_{n+n+} + \bm{K}_{n-n-}) \bm{L}_p^{T},
\end{equation}\end{linenomath*}
where $\bm{L}_p$ is given in
equation~(\ref{eqn:jacobian:constraint:code}) and $\bm{K}_{n+n+}$
and $\bm{K}_{n-n-}$ are the diagonal terms from
equation~(\ref{eqn:saddle:point:code}).


Our preferred setup uses the field splitting options in PETSc to
combine an AMG preconditioner for the elasticity submatrix with our
custom fault preconditioner for the Lagrange multiplier submatrix. See
section~\ref{sec:performance:benchmark} for a comparison of
preconditioner performance for an application involving a static
simulation with multiple faults. It shows the clear superiority of
this setup over several other possible preconditioning strategies.

\subsection{Dynamic Simulations}

In dynamic simulations the Courant-Friderichs-Lewy condition
\citep{Courant:etal:1967} controls the stability of the explicit time
integration. In most dynamic problems this dictates a relatively small
time step so that a typical simulation involves tens of thousands of
time steps. Hence, we want a very efficient solver to run
dynamic simulations in a reasonable amount of time.


The Jacobian for our system of equations involves two terms: the
inertial term given by equation~(\ref{eqn:jacobian:explicit:inertia})
and the fault slip constraint term given by
equation~(\ref{eqn:jacobian:constraint}). Using conventional
finite-element basis functions in these integrations results in a
sparse matrix with off-diagonal terms. Although we can use the same
solvers as we do for quasi-static simulations to find the solution,
eliminating the off-diagonal terms so that the Jacobian is diagonal
permits use of a much faster solver. With a diagonal Jacobian the
number of operations required for the solve is proportional to the
number of DOF, and the memory requirements are greatly
reduced by storing the diagonal of the matrix as a vector rather
than as a sparse matrix. However, the block structure of our Jacobian
matrix, with the fault slip constraints occupying off-diagonal blocks,
requires a two step approach to solve the linear system
of equations without forming a sparse matrix.

First, we eliminate the off-diagonal entries in each block of the
matrix during the finite-element integrations.  The current best
available option for eliminating the off-diagonal terms formed during
the integration of the inertial term focuses on choosing a set of
orthogonal basis functions, such as the Legendre polynomials with
Gauss-Lobatto-Legendre quadrature points
\citep{Komatitsch:Vilotte:1998}. This discretization (often called the
spectral element method) naturally produces a diagonal block for each
finite-element cell without introducing any additional
approximations. Because the fault slip constraint term also involves
integration of the products of the basis functions over
lower-dimension cells, orthogonal basis functions also produce a
diagonal block for this integration.

In contrast, traditional finite-element approaches do introduce
additional approximations when constructing a diagonal
approximation. In PyLith we employ one of these traditional
approaches, because it produces good approximations for many different
choices of basis functions and quadrature points. For each
finite-element cell, we construct a diagonal approximation of the
integral such that the action on rigid body motion is the same for the
diagonal approximation of the integral as it is for the original
integral,
\begin{linenomath*}\begin{equation}
  \bm{A} \cdot \bm{u}_\mathrm{rigid} =
  \bm{A}_\mathit{diagonal} \cdot \bm{u}_\mathrm{rigid}.
\end{equation}\end{linenomath*}
Expressing the diagonal block of the Jacobian matrix as a vector and
the matrix of basis functions as a vector we have,
\begin{linenomath*}\begin{equation}
  \bm{A}  = \int_\Omega \bm{N}^T \cdot \bm{N} \, d\Omega \rightarrow
  \bm{A}_\mathit{diagonal} = \int_\Omega \bm{N} \sum_i N_i \, d\Omega,
\end{equation}\end{linenomath*}
where $N_i$ is the scalar basis function for degree of freedom $i$ and
$\Omega$ may be the domain volume (as in the case of the inertial
term) or a boundary (as in the case of the fault slip constraint
term).

The errors associated with this approximation are small as
long as the deformation occurs at length scales significantly larger
than the discretization size, which is consistent with resolving
seismic wave propagation accurately. Furthermore, in contrast to other
approaches that choose basis functions or quadrature points that
affect the accuracy of all of the finite-element integrations, such as
choosing quadrature points coincident with the vertices of a cell,
this approach only affects the accuracy of the terms involved in the
Jacobian. For consistency in the formulation of the system of
equations, these approximations are also applied to the inertial term
and fault slip constraint term when computing the residual.


Second, we leverage the structure of the off-diagonal blocks
associated with the fault slip constraint in solving the system of
equations via a Schur's complement algorithm. We compute an initial
residual assuming the increment in the solution is zero (i.e.,
$\bm{du}_n = \bm{0}$ and $\bm{dl}_p = \bm{0}$),
\begin{linenomath*}\begin{equation}
  \bm{r}^* = \begin{pmatrix} \bm{r}_n^* \\ \bm{r}_p^* \end{pmatrix} =
  \begin{pmatrix} \bm{b}_n \\ \bm{b}_p \end{pmatrix}
  - \begin{pmatrix}
    \bm{K} & \bm{L}^T \\ \bm{L} & 0
  \end{pmatrix}
  \begin{pmatrix} \bm{u}_n \\ \bm{l}_p \end{pmatrix}.
\end{equation}\end{linenomath*}
 We compute a corresponding initial solution to the system of equations
$\bm{du}_n^*$ ignoring the off-diagonal blocks in the Jacobian and
the increment in the Lagrange multipliers.
\begin{linenomath*}\begin{equation}
\bm{du}_n^* = \bm{K}^{-1} \cdot \bm{r}_n,
\end{equation}\end{linenomath*}
taking advantage of the fact that we construct $\bm{K}$ so that it
is diagonal. 

We next compute the increment in the Lagrange multipliers to
correct this initial solution so that the true residual is zero.
Making use of the initial residual, the expression for the true
residual is
\begin{linenomath*}\begin{equation}
  \label{eqn:lumped:jacobian:residual}
  \bm{r} = \begin{pmatrix} \bm{r}_n \\ \bm{r}_p \end{pmatrix} =
  \begin{pmatrix} \bm{r}_n^* \\ \bm{r}_p^* \end{pmatrix}
  - \begin{pmatrix}
    \bm{K} & \bm{L}^T \\ \bm{L} & 0
  \end{pmatrix}
  \begin{pmatrix} \bm{du}_n \\ \bm{dl}_p \end{pmatrix}.
\end{equation}\end{linenomath*}
Solving the first row of equation~(\ref{eqn:lumped:jacobian:residual})
for the increment in the solution and accounting for the structure of
$\bm{L}$ as we write the expressions for DOF on
each side of the fault, we have
\begin{linenomath*}\begin{gather}
  \bm{du}_{n^+} = 
    \bm{du}_{n^+}^* - \bm{K}_{n^+n^+}^{-1} \cdot \bm{L}_p^T \cdot \bm{dl}_p, \\
  \bm{du}_{n^-} = 
    \bm{du}_{n^-}^* + \bm{K}_{n^-n^-}^{-1} \cdot \bm{L}_p^T \cdot \bm{dl}_p.
\end{gather}\end{linenomath*}
Substituting into the second row of
equation~(\ref{eqn:lumped:jacobian:residual}) and isolating the term
with the increment in the Lagrange multipliers yields
\begin{linenomath*}\begin{equation}
  \bm{L}_p \cdot 
  \left( \bm{K}_{n^+n^+}^{-1} + \bm{K}_{n^-n^-}^{-1} \right) \cdot 
  \bm{L}^T_p \cdot \bm{dl}_p =
  -\bm{r}_p^* + \bm{L}_p \cdot 
  \left( \bm{du}_{n^+}^* - \bm{du}_{n^-}^* \right).
\end{equation}\end{linenomath*}
Letting
\begin{linenomath*}\begin{equation}
  \bm{S}_p = \bm{L}_p \cdot 
  \left( \bm{K}_{n^+n^+}^{-1} + \bm{K}_{n^-n^-}^{-1} \right) \cdot 
  \bm{L}^T_p,
\end{equation}\end{linenomath*}
and recognizing that $\bm{S}_p$ is diagonal because $\bm{K}$
and $\bm{L}_p$ are diagonal allows us to solve for the increment
in the Lagrange multipliers,
\begin{linenomath*}\begin{equation}
  \bm{dl}_p = \bm{S}_p^{-1} \cdot \left[
  -\bm{r}_p^* + \bm{L}_p \cdot 
  \left( \bm{du}_{n^+}^* - \bm{du}_{n^-}^* \right)
  \right].
\end{equation}\end{linenomath*}
Now that we have the increment in the Lagrange multipliers, we can
correct our initial solution $\bm{du}_n^*$ so that the true residual
is zero,
\begin{linenomath*}\begin{equation}
  \bm{du}_n = 
  \bm{du}_n^* - \bm{K}^{-1} \cdot \bm{L}^T \cdot \bm{dl}_p.
\end{equation}\end{linenomath*}
Because $\bm{K}$ and $\bm{L}$ are comprised of diagonal
blocks, this expression for the updates to the solution are local to
the DOF attached to the fault and the Lagrange
multipliers.


We also leverage the elimination of off-diagonal entries from the
blocks of the Jacobian in dynamic simulations when updating the slip
in spontaneous rupture models. Because $\bm{K}$ is diagonal in
this case, the expression for the change in slip for a perturbation in
the Lagrange multipliers
(equations~(\ref{eqn:spontaneous:rupture:update:lagrange})--(\ref{eqn:spontaneous:rupture:update:slip}))
simplifies to
\begin{linenomath*}\begin{equation}
  \partial \bm{d}_p = - \left( \bm{K}_{n^+n^+}^{-1} + \bm{K}_{n^-n^-}^{-1} \right)
  \cdot \bm{L}_p^T \cdot \partial \bm{l}_p.
\end{equation}\end{linenomath*}
Consequently, the increment in fault slip and Lagrange multipliers for
each vertex can be done independently. In dynamic simulations the time
step is small enough that the fault constitutive model is much less
sensitive to the slip than in most quasi-static simulations, so we 
avoid performing a line search in computing the update.

\section{Performance Benchmark}
\label{sec:performance:benchmark}

\begin{figure}
  \noindent\includegraphics{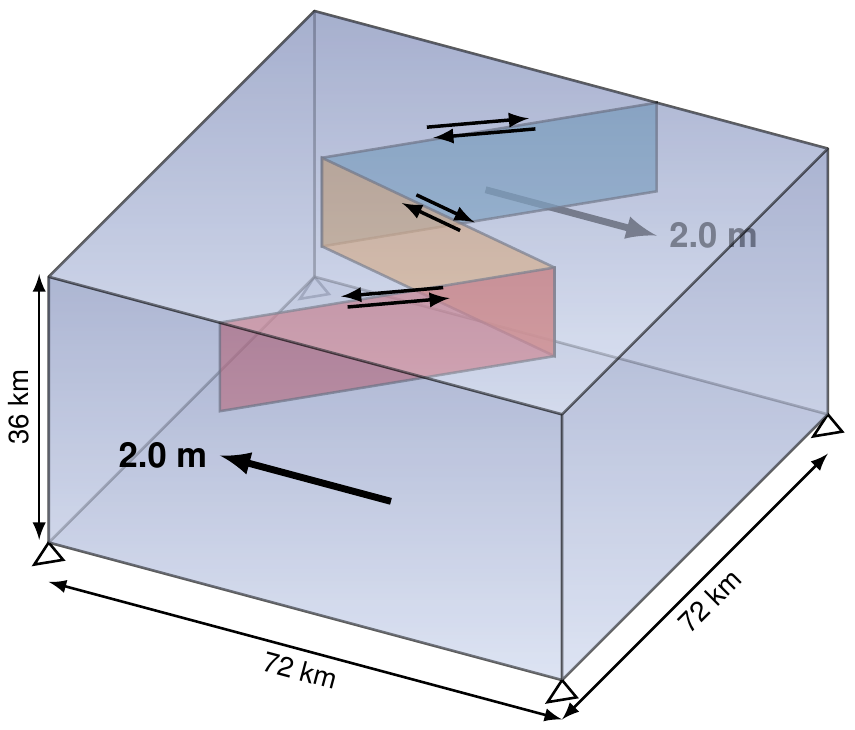}
  \caption{Geometry of problem used in quasi-static performance
    benchmark. Dirichlet boundary conditions prescribe a horizontal
    lateral displacement of 2.0 m with no motion normal to the
    boundary on two sides of the domain and zero vertical displacement
    on the bottom boundary. We specify uniform slip of 1.0 m of
    right-lateral motion on the middle fault and 0.5 m of left-lateral
    motion on the two other faults. The faults extend down to a depth
    of 12.0 km.}
  \label{fig:solvertest:geometry}
\end{figure}

We compare the relative performance of the various preconditioners
discussed in section~\ref{sec:solver:quasi-static} for quasi-static
problems using a static simulation with three vertical, strike-slip
faults. Using multiple, intersecting faults introduces multiple saddle
points, so it provides a more thorough test of the preconditioner
compared to a single fault with a single saddle point.
Figure~\ref{fig:solvertest:geometry} shows the geometry of the faults
embedded in the domain and Table~\ref{tab:solvertest:parameters} gives
the parameters used in the simulation. We apply Dirichlet boundary
conditions on two lateral sides with 2.0 m of shearing motion and no
motion perpendicular to the boundary. We also apply a Dirichlet
boundary condition to the bottom of the domain to prevent vertical
motion. We prescribe uniform slip on the three faults with zero slip
along the buried edges.

\begin{figure}
  \noindent\includegraphics[width=84mm]{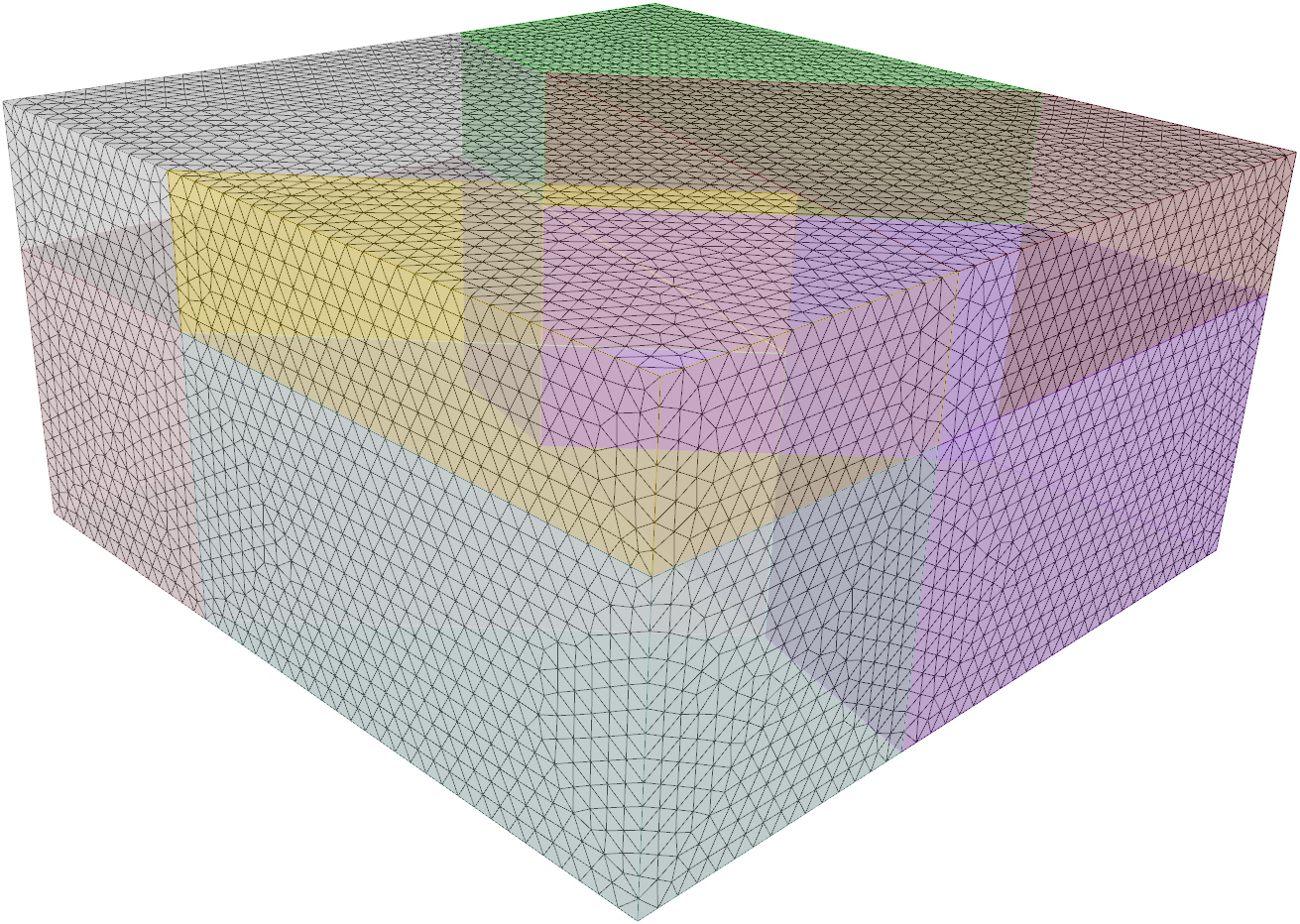}
  \caption{Tetrahedral finite-element mesh with a uniform
    discretization size of 1744 m for the performance benchmark. The
    colors correspond to the volumes in the CUBIT geometry that are
    separated by the fault surfaces and boundary between the upper
    crust and lower crust.}
  \label{fig:solvertest:mesh}
\end{figure}

We generate both hexahedral meshes and tetrahedral meshes using CUBIT
(available from \url{http://cubit.sandia.gov}) and construct meshes so that
the problem size (number of DOF) for the two different cell types
(hexahedra and tetrahedra) are nearly the same (within 2\%). The suite
of simulations examines increasingly larger problem sizes as we increase
the number of processes (with one process per core), with $7.8\times
10^4$ DOF for 1 process up to $7.1\times 10^6$ DOF for 96
processes. The corresponding discretization sizes are 2033 m to 437 m
for the hexahedral meshes and 2326 m to 712 m for the tetrahedral
meshes.  Figure~\ref{fig:solvertest:mesh} shows the 1846 m resolution
tetrahedral mesh. As we will see in
section~\ref{sec:verification:quasi-static}, the hexahedral mesh for a
given resolution in a quasi-static problem is slightly more accurate,
so the errors in solution for each pair of meshes are larger for the
tetrahedral mesh.

\subsection{Preconditioner Performance}

We characterize preconditioner performance in terms of the number of
iterations required for the residual to reach a given convergence
tolerance and the sensitivity of the number of iterations to the
problem size. Of course, we also seek a minimal overall computation
time. We examine the computation time in the next section when
discussing the parallel performance. An ideal preconditioner would
yield a small, constant number of iterations independent of problem
size. However, for complex problems such as elasticity with fault slip
and potentially nonuniform physical properties, ideal preconditioners
may not exist. Hence, we seek a preconditioner that provides a minimal
increase in the number of iterations as the problem size increases, so
that we can efficiently simulate quasi-static crustal deformation
related to faulting and post-seismic and interseismic deformation.

For this benchmark of preconditioner performance, we examine the
number of iterations required for convergence using the PETSc additive
Schwarz (ASM), field split (with and without our custom
preconditioner), and Schur complement preconditioners discussed in
section~\ref{sec:solver:quasi-static}. We characterize the dependence
on problem size using serial simulations (we examine parallel scaling
for the best preconditioner in the next section) and the three lowest
resolution meshes in our suite of hexahedral and tetrahedral meshes
with the results summarized in
Table~\ref{tab:solvertest:preconditioner:iterates}.

The Schur complement and family of field split preconditioners using
algebraic multigrid methods minimize the increase in the number of
iterations with problem size. For these preconditioners the number of
iterations increases by only about 20\% for a four times increase in
the number of degrees of freedom, compared to 60\% for the ASM
preconditioner. Within the family of field split preconditioners using
algebraic multigrid methods, the one with multiplicative composition
minimizes the number of iterations. The custom preconditioner for the
Lagrange multiplier submatrix greatly accelerates the convergence with
an 80\% reduction in the number of iterations required for
convergence. This preconditioner also provides the fastest runtime of
all of these preconditioners.

\subsection{Parallel Scaling Performance}

The underlying PETSc solver infrastructure has demonstrated optimal
scalability on the largest machines available today
\citep{srksyp2008,GordonBell09,msmhls2010,brown2012tmeice}. However,
computer science scalability results are often based upon
unrealistically simple problems which do not advance the scientific
state-of-the-art. In evaluating the parallel scalability of PyLith, we
consider the sources responsible for reducing the scalability and
propose possible steps for mitigation.

The main impediment to scalability in PyLith is load imbalance in
solving the linear system of equations.  This imbalance is the
combination of three effects: the inherent imbalance in partitioning
an unstructured mesh, partitioning based on cells rather than DOF, and
weighting the cohesive cells the same as conventional bulk cells while
partitioning. In this performance benchmark matrix-vector
multiplication (the PETSc \texttt{MatMult} function) has a load
imbalance of up to 20\% on 96 cores.  The cell partition balances
the number of cells across the processes using ParMetis
\citep{Karypis:etal:1999} to achieve good balance for the
finite element integration. This does not take into account a
reduction in the number of DOF associated with constraints from
Dirichlet boundary conditions or the additional DOF associated with
the Lagrange multiplier constraints, which can exacerbate any
imbalance. Nevertheless, eliminating DOF associated with Dirichlet
boundary conditions preserves the symmetry of the overall systems and,
in many cases, results in better conditioned linear systems.

We evaluate the parallel performance via a weak scaling
criterion. That is, we run simulations on various numbers of
processors/cores with an increase in the problem size as the number of
processes increases (with one process per core) to maintain the same
workload (e.g., number of cells and number of DOF) for each core. In
ideal weak scaling the time for the various stages of the simulation
is independent of the number of processes. For this performance
benchmark we use the entire suite of hexahedral and tetrahedral meshes
described earlier that range in size from $7.8\times 10^4$ DOF (1
process) to $7.1\times 10^6$ DOF (96 processes). We employ the AMG
preconditioner for the elasticity submatrix and our custom
preconditioner for the Lagrange multipliers submatrix. We ran the
simulations on Lonestar at the Texas Advanced Computing
Center. Lonestar is comprised of 1888 compute nodes connected by QDR
Infiniband in a fat-tree topology, where each compute node consists
of two six-core Intel Xeon E5650 processors with 24 GB of
RAM. Simulations run on twelve or fewer cores were run on a single
compute node with processes distributed across processors and then
cores. For example, the two process simulation used one core on each
of two processors. In addition to algorithm bottlenecks, runtime
performance is potentially impeded by core/memory affinity, memory
bandwidth, communication among compute nodes (including communication
from other jobs running on the machine).

The single node scaling for PyLith (twelve processes or less in this
case) is almost completely controlled by the available memory
bandwidth. Good illustrations of the memory system performance are
given by the \texttt{VecAXPY}, \texttt{VecMAXPY} and \texttt{VecMDot}
operations reported in the log summary \citep{PETSc:manual}. These
operations are limited by available memory bandwidth rather than the
rate at which a processor or core can perform floating points operations. From
Table~\ref{tab:solvertest:memory:events}, we see that we saturate the
memory bandwidth using two processes (cores) per processor, since scaling
plateaus from 2 to 4 processes, but shows good scaling from 12 to 24
processes. This lack of memory bandwidth will depress overall
performance, but should not affect the inter-node scaling of the
application.

Machine network performance can be elucidated by the \texttt{VecMDot}
operation for vector reductions, and \texttt{MatMult} for
point-to-point communication. In
Table~\ref{tab:solvertest:memory:events} we see that the vector
reduction shows good scaling up to 96 processes. Similarly in
Table~\ref{tab:solvertest:solver:events}, we see that \texttt{MatMult}
has good scalability, but that it is a small fraction of the overall
solver time. The AMG preconditioner setup (\texttt{PCSetUp}) and
application (\texttt{PCApply}) dominate the overall solver time. The
AMG preconditioner setup time increases with the number of
processes. Note that many weak scaling studies do not include this
event, because it is amortized over the iteration. Nevertheless, in
our benchmark it is responsible for most of the deviation from perfect
weak scaling.  We could trade preconditioner strength for scalability
by reducing the work done on the coarse AMG grids, so that the solver
uses more iterations which scale very well.  However, that would
increase overall solver time and thus would not be the choice to
maximize scientific output.

\begin{figure}
  \noindent\includegraphics{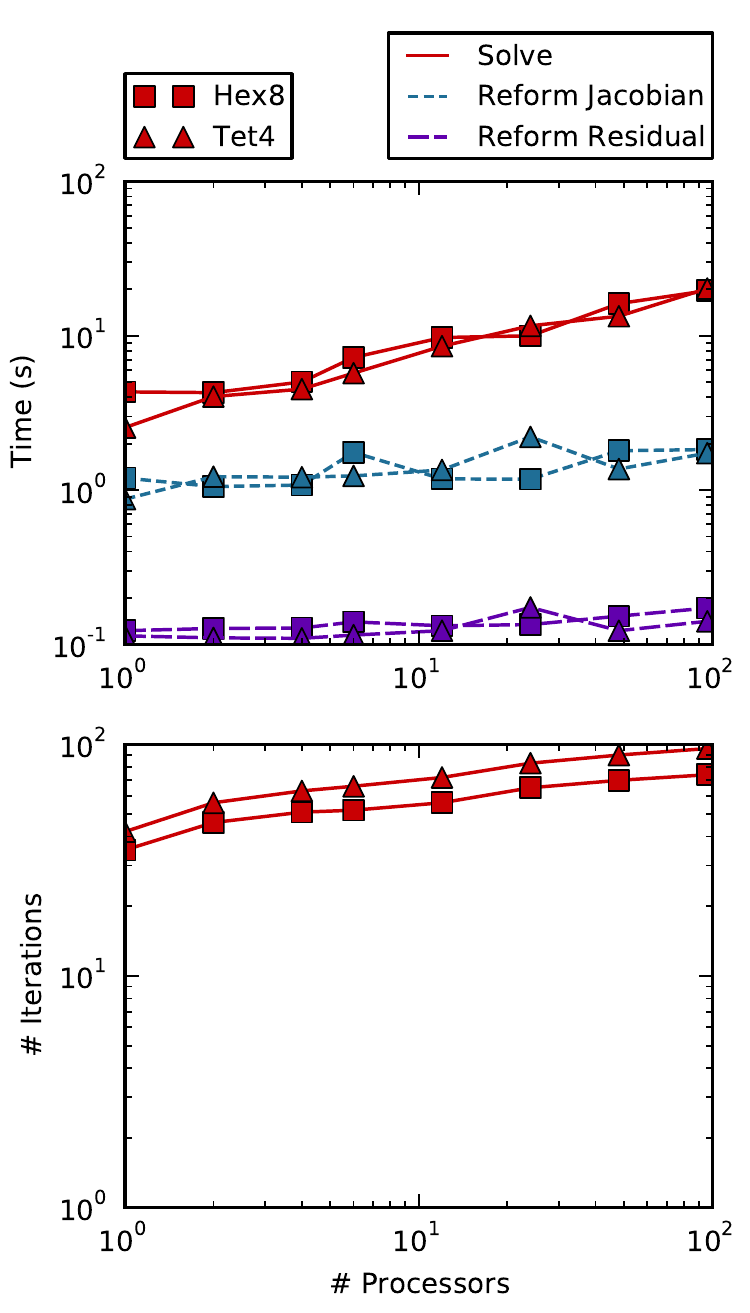}
  \caption{Plot of parallel scaling for the performance benchmark with
    the algebraic multigrid preconditioner and fault block custom
    preconditioner. The stages shown include the numerical integration
    of the residual ({\tt Reform Residual}) and Jacobian ({\tt Reform
      Jacobian}) and setting up the preconditioner and solving the
    linear system of equations ({\tt Solve}). The finite-element
    integrations for the Jacobian and residual exhibit good weak
    scaling with minimal sensitivity to the problem size. The linear
    solve (solid lines in the top panel) does not scale as well, which
    we attribute to the poor scaling of the algebraic multigrid setup
    and application as well as limited memory and interconnect
    bandwidth. We attribute fluctuations in the relative performance
    to variations in the machine load from other jobs on the cluster.}
  \label{fig:solvertest:scaling}
\end{figure}

Figure~\ref{fig:solvertest:scaling} illustrates the excellent parallel
performance for the finite-element assembly routines (reforming the
Jacobian sparse matrix and computing the residual). As discussed
earlier in this section, the ASM preconditioner performance is not
scalable because the number of iterations increases significantly with
the number of processes. As shown in
Figure~\ref{fig:solvertest:scaling}, the introduction of Schur
complement methods and an AMG preconditioner slows the growth
considerably, and future work will pursue the ultimate goal of
iteration counts independent of the number of processes.

\section{Code Verification Benchmarks}
\label{sec:verification:benchmarks}

In developing PyLith we verify the numerical implementation of various
features using a number of techniques. We employ unit testing to
verify correct implementation of nearly all of the individual
routines. Having a test for most object methods or functions isolates
bugs at their origin during code development and prevents new bugs
from occurring as code is modified or optimized. We also rely on
full-scale benchmarks to verify that the code properly solves the
numerical problem.  These benchmarks include quasi-static strike-slip
and reverse viscoelastic simulations and various exercises in the
suite of dynamic spontaneous rupture benchmarks developed by the
Southern California Earthquake Center (SCEC) and the United States
Geological Survey \citep{Harris:etal:SRL:2009}. The mesh generation
and simulation parameter files for many of the benchmarks, including
those discussed here, are available from the CIG subversion repository
(\url{http://geodynamics.org/svn/cig/short/3D/PyLith/benchmarks/trunk/}). In
this section we focus on two benchmarks that test different scientific
applications: quasi-static relaxation of a Maxwell viscoelastic
material subjected to multiple earthquake cycles involving slip and
steady creep on a vertical strike-slip fault
\citep{Savage:Prescott:1978} and supershear dynamic spontaneous
rupture of a 60 degree dipping normal fault in a Drucker-Prager
elastoplastic medium. This second benchmark corresponds to benchmark
TPV13 in the SCEC suite of dynamic spontaneous rupture benchmarks
\citep{Harris:etal:SRL:2011}.

\subsection{Quasi-static Benchmark}
\label{sec:verification:quasi-static}

\begin{figure}
  \noindent\includegraphics[width=84mm]{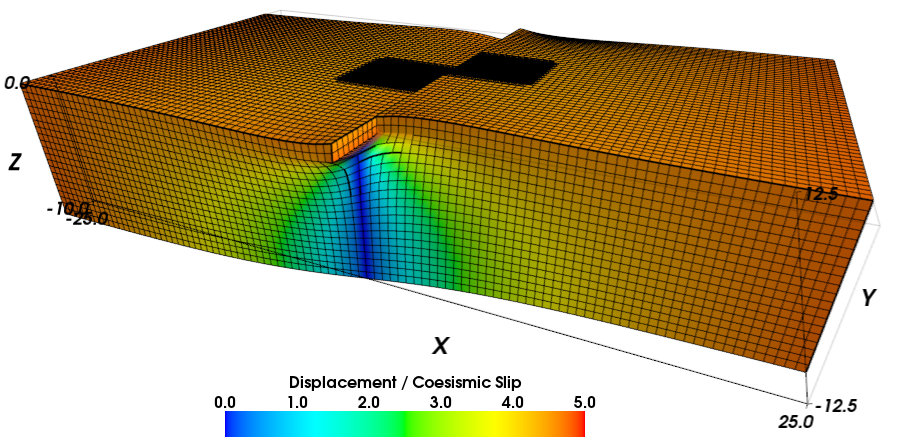}
  \caption{Deformation (exaggerated by a factor of 5000) 95\% of the
    way through earthquake cycle 10 of the Savage and Prescott
    benchmark, which involves viscoelastic relaxation over multiple
    earthquake cycles on a vertical, strike-slip fault. The
    coordinates are in units of elastic layer thickness and the
    displacement field is in units of coseismic slip. The locking
    depth is one-half of the thickness of the elastic layer. We refine
    the hexahedral mesh by a factor of three near the center of the
    domain. Figure~\ref{fig:savage:prescott:profiles} compares
    profiles along y=0 with the analytic solution.}
  \label{fig:savage:prescott::solution}
\end{figure}

As a test of our quasi-static solution, we compare our numerical
results against the analytical solution of
\citet{Savage:Prescott:1978}. This problem consists of an infinitely
long strike-slip fault in an elastic layer overlying a Maxwell
viscoelastic half-space. The parameter files for this benchmark are
available in the {\tt quasistatic/sceccrustdeform/savageprescott} directory
of the benchmark repository. Figure~\ref{fig:savage:prescott::solution}
illustrates the geometry of the problem with an exaggerated view of
the deformation during the tenth earthquake cycle. Between earthquakes
the upper portion of the fault is locked, while the lower portion
slips at the plate velocity. At regular intervals (the earthquake
recurrence time) the upper portion of the fault slides such that the
slip on the locked portion exactly complements the slip on the
creeping portion so the cumulative slip over an earthquake cycle is
uniform.

This problem tests the ability of the kinematic fault implementation to
include steady aseismic creep and multiple earthquake ruptures 
along with viscoelastic relaxation. The analytical solution for this
problem provides the along-strike component of surface displacement as
a function of distance perpendicular to the fault. The solution is
controlled by the ratio of the fault locking depth to the thickness of
the elastic layer and the ratio of the earthquake recurrence time to
the viscoelastic relaxation time, $\tau_0 = \mu T/2\eta$, where $T$ is
the recurrence time, $\mu$ is the shear modulus and $\eta$ is the
viscosity.

For this benchmark we use a locking depth of 20 km, an elastic layer
thickness of 40 km, an earthquake recurrence time of 200 years, a
shear modulus of 30 GPa, a viscosity of $2.37\times 10^{19}$ Pa-s, and
a relative plate velocity of 2 cm/year, implying a coseismic offset of
4 m every 200 years (see
Table~\ref{tab:Savage:Prescott:parameters}). The viscosity and shear
modulus values yield a viscoelastic relaxation time of 50 years, and
$\tau_0 =4$.  We employ a 3-D model (2000 km by 1000 km by 400 km)
with Dirichlet boundary conditions enforcing symmetry to approximate
an infinitely long strike-slip fault. We apply velocity boundary
conditions in the y-direction to the -x and +x faces with zero
x-displacement. We constrain the vertical displacements on the bottom
of the domain to be zero. Finally, we fix the x-displacements on the
-y and +y faces to enforce symmetry consistent with an infinitely long
strike-slip fault.

We examine four different numerical solutions considering the effects
of cell type (hexahedral versus tetrahedral) and discretization
size. In our coarse hexahedral mesh we use a uniform resolution of 20
km. In our higher resolution hexahedral mesh we refine an inner region
(480 km by 240 km by 100 km) by a factor of three, yielding a
resolution near the center of the fault of 6.7 km. For the
tetrahedral meshes, we match the discretization size of the hexahedral
mesh near the center of the fault (20km or 6.7 km) while increasing
the discretization size in a geometric progression at a rate of
1.02. This results in a maximum discretization size of approximately
60 km for the coarser mesh and 40 km for the higher resolution
mesh. Note that for both the hexahedral and tetrahedral coarse meshes,
the discretization size on the fault is the maximum allowable size
that still allows us to represent the fault locking depth as a sharp
boundary.

In this viscoelastic problem neither the analytical or numerical
models approach steady-state behavior until after several earthquake
cycles. There is also a difference in how steady plate motion is
applied for the two models. For the analytical solution, steady plate
motion is simply superimposed, while for the numerical solution steady
plate motion is approached after several earthquake cycles, once the
applied fault slip and velocity boundary conditions have produced
nearly steady flow in the viscoelastic half-space. It is therefore
necessary to spin-up both solutions to their steady-state solution
over several earthquake cycles to allow a comparison between the
two. In this way, the transient behavior present in both models will
have nearly disappeared, and both models will have approximately the
same component of steady plate motion. We simulate ten earthquake
cycles for both the analytical and numerical models for a total
duration of 2000 years. For the numerical solution we use a constant
time step size of five years. This time step corresponds to one tenth
of the viscoelastic relaxation time; hence it tests the accuracy of
the viscoelastic solution for moderately large time steps relative to
the relaxation time. Recall that the quasi-static formulation does not
include inertial terms and time stepping is done via a series of
static problems so that the temporal accuracy depends only on the
temporal variation of the boundary conditions and constitutive models.
These benchmarks simulations can be run on a laptop or desktop
computer. For example, the high resolution benchmarks took 46 min
(hexahedral cells) and 36 min (tetrahedral cells) using four processes
on a dual quad core desktop computer with Intel Xeon E5630 processors.

\begin{figure}
  \noindent\includegraphics{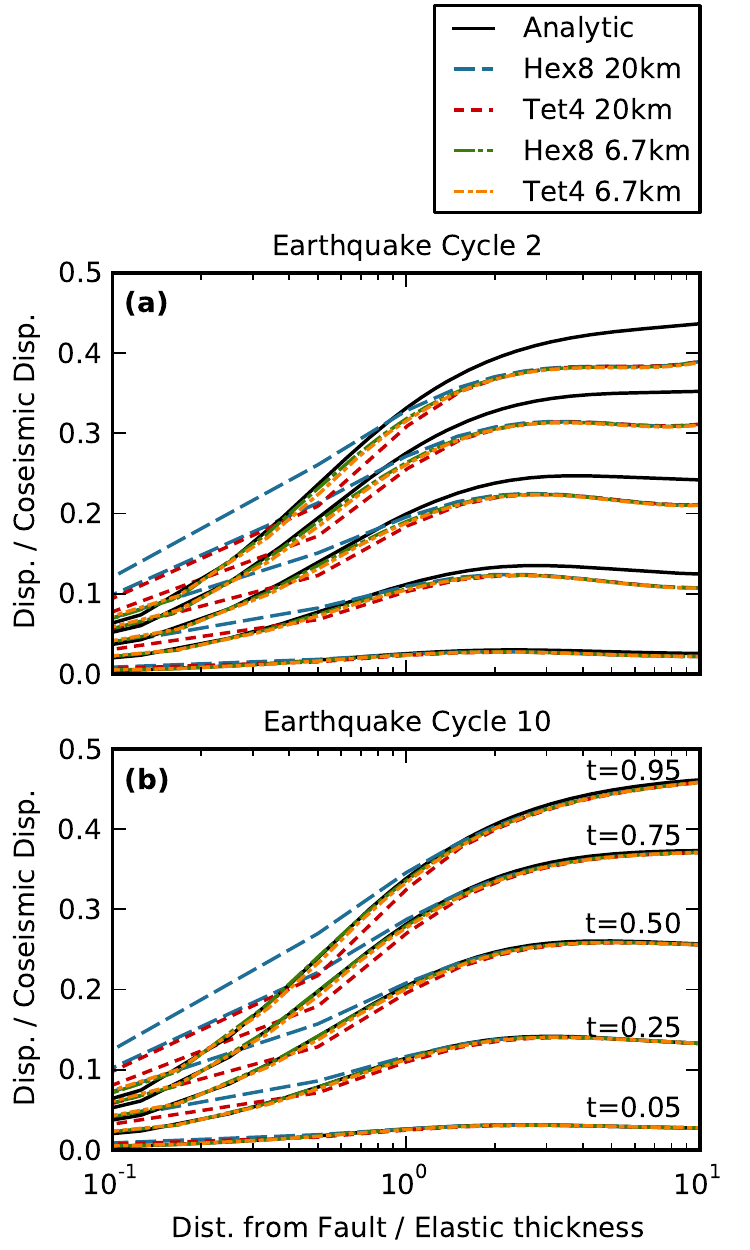}
  \caption{Comparison of displacement profiles perpendicular to the
    fault in the Savage and Prescott benchmark during earthquake
    cycles (a) two and (b) ten. The displacement values shown are
    relative to the values at the beginning of the earthquake cycle to
    facilitate comparison between the analytical solution and the
    numerical models. Both the analytical and numerical simulations
    require spin-up to reach the steady-state solution, and the
    numerical models also require spin-up to achieve steady plate
    motion, which is superimposed on the analytical solution. Both the
    hexahedral (Hex8) and tetrahedral (Tet4) discretizations resolve
    the viscoelastic deformation and display excellent agreement with
    the steady-state solution by the tenth earthquake cycle. The
    coarser (20 km) resolutions are unable to match the details of the
    displacement field at distances less than about one elastic
    thickness, but all of the numerical models provide a good fit to
    the analytical solution in the tenth earthquake cycle at distances
    greater than 2--3 times the elastic thickness.}
  \label{fig:savage:prescott:profiles}
\end{figure}

Figure~\ref{fig:savage:prescott:profiles} compares the numerical
results extracted on the ground surface along the center of the model
perpendicular to the fault with the analytic solution. Using a
logarithmic scale with distance from the fault facilitates examining
the solution both close to and far from the fault. For the second
earthquake cycle, the far-field numerical solution does not yet
accurately represent steady plate motion and the numerical simulations
underpredict the displacement. By the tenth earthquake cycle, steady
plate motion is accurately simulated and the numerical results match
the analytical solution.

Within about one elastic thickness of the fault the effect of the
resolution of the numerical models becomes apparent. We find large
errors for the coarse models, which have discretization sizes matching
the fault locking depth. The finer resolution models (6.7 km
discretization size) provide a close fit to the analytical
solution. The 6.7 km hexahedral solution is indistinguishable from the
analytical solution in Figure~\ref{fig:savage:prescott:profiles}(b);
the 6.7 km tetrahedral solution slightly underpredicts the analytical
solution for times late in the earthquake cycle.  The greater
accuracy of the hexahedral cells relative to the tetrahedral cells
with the same nominal discretization size for quasi-static solutions
is consistent with our findings for other benchmarks. The greater
number of polynomial terms in the basis functions of the hexahedra
allows the model to capture a more complex deformation field at a
given discretization size.

\subsection{Dynamic Benchmark}
\label{sec:verification:dynamic}

As a test of PyLith's dynamic spontaneous rupture solutions, we use
SCEC Spontaneous Rupture Benchmark TPV13 that models a high
stress-drop, supershear, dip-slip earthquake that produces extreme
(very large) ground motions, large slip, and fast slip rates
\citep{Harris:etal:SRL:2011}. It uses a Drucker-Prager elastoplastic
bulk rheology and a slip-weakening friction model in a depth-dependent
initial stress field. The parameter files for this benchmark are
available in the {\tt dynamic/scecdynrup/tpv210-2d} and
{\tt dynamic/scecdynrup/tpv210} directories of the benchmark repository.

\begin{figure}
  \noindent\includegraphics{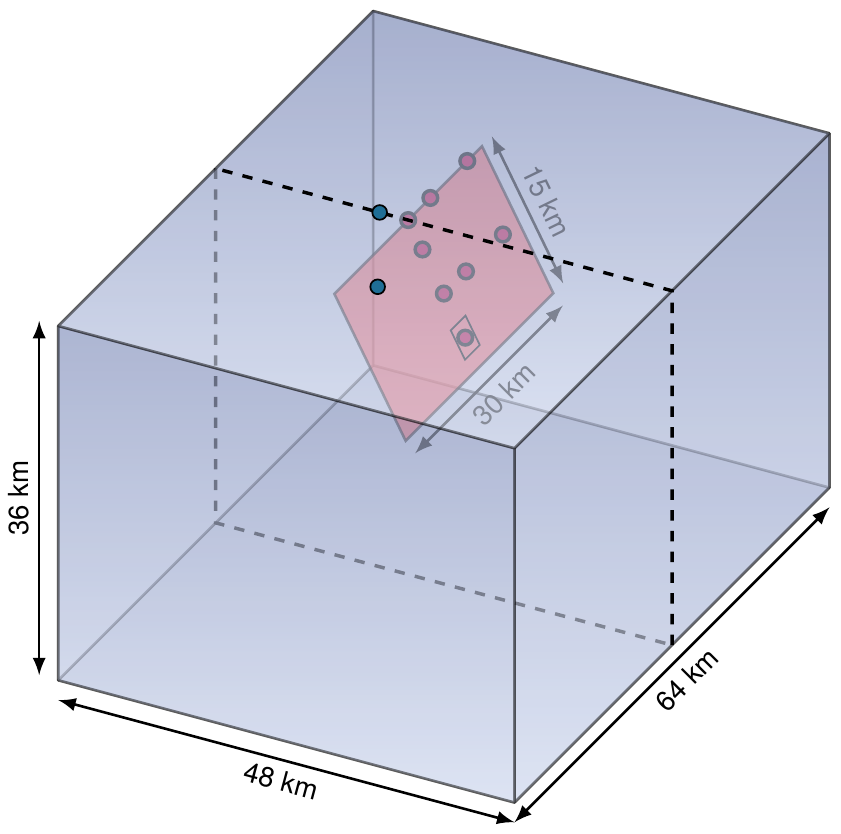}
  \caption{Geometry for SCEC spontaneous rupture benchmark TPV13 involving
    a Drucker-Prager elastoplastic bulk rheology, slip-weakening
    friction, a depth-dependent stress field, and normal fault with a
    60 degree dip angle. The 2-D version corresponds to the vertical
    slice shown by the dashed line. The red dots denote locations on
    the fault used in the comparison of the vertical slip rates
    (Figures~\ref{fig:tpv13-2d:slip:rate}
    and~\ref{fig:tpv13:slip:rate}). The blue dots indicate locations
    on the ground surface used in the comparison of fault normal and
    vertical velocity time histories (Figure~\ref{fig:tpv13:velocity}).}
  \label{fig:tpv13:geometry}
\end{figure}

\begin{figure}
  \noindent\includegraphics[width=84mm]{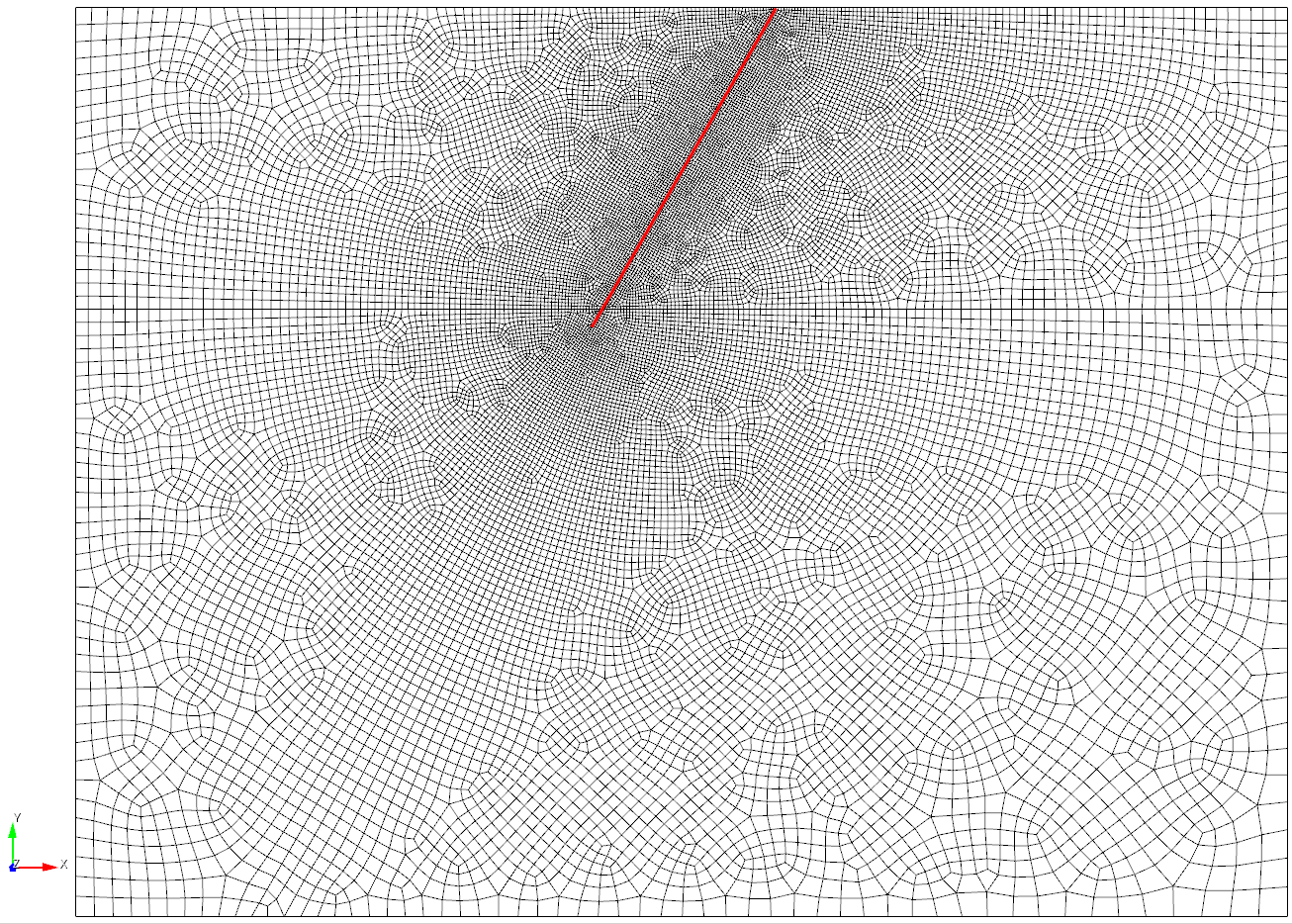}
  \caption{Finite-element mesh comprised of quadrilateral cells for SCEC
    spontaneous rupture benchmark TPV13-2D. The discretization size is 100
    m on the fault surface and increases at a geometric rate of 2\%
    with distance from the fault. We employ this same spatial
    variation of the discretization size in the 3-D model.}
  \label{fig:tpv13-2d:mesh}
\end{figure}

Figure~\ref{fig:tpv13:geometry}
shows the geometry of the benchmark and the size of the domain
we used in our verification test. The benchmark includes both 2-D
(TPV13-2D is a vertical slice through the fault center-line with plane
strain conditions) and 3-D versions (TPV13). This benchmark specifies
a spatial resolution of 100 m on the fault surface. To
examine the effects of cell type and discretization size we consider
both triangular and quadrilateral discretizations with resolutions on
the fault of 50 m, 100 m, and 200 m for TPV13-2D and 100 m and 200 m
for TPV13. We gradually coarsen the mesh with distance from the fault
by increasing the discretization size at a geometric rate of 2\%. This
provides high resolution at the fault surface to resolve the small
scale features of the rupture process and less resolution at the
edges of the boundary where the solution is much
smoother. Figure~\ref{fig:tpv13-2d:mesh} shows the triangular mesh for
a discretization size of 100 m on the fault.

\begin{figure}
  \noindent\includegraphics{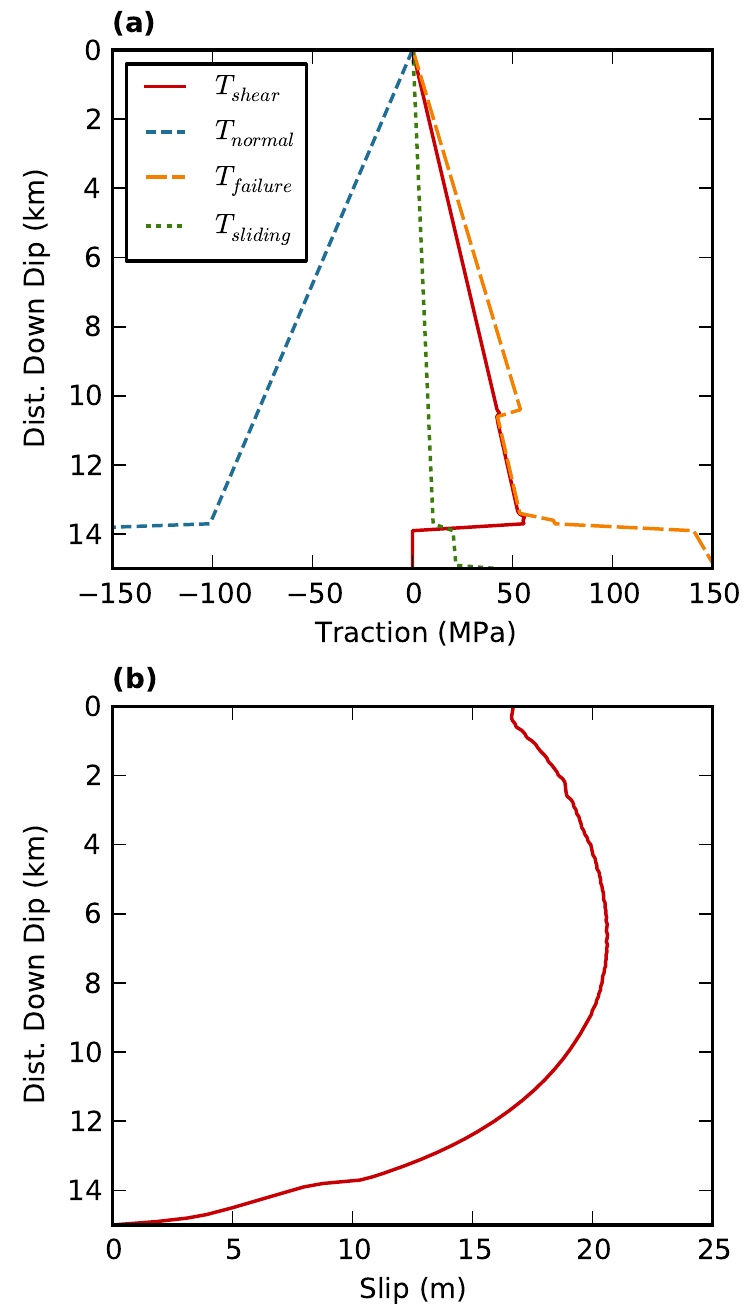}
  \caption{(a) Depth-dependent fault tractions in SCEC spontaneous rupture
    benchmark TPV13-2D and TPV13. $T_\mathit{shear}$ denotes the
    initial shear traction, $T_\mathit{normal}$ denotes the initial
    effective normal traction, $T_\mathit{failure}$ denotes the
    frictional failure stress corresponding to the initial effective
    normal traction, and $T_\mathit{sliding}$ denotes the dynamic
    sliding stress corresponding to the initial effective normal
    traction. Positive shear tractions correspond to normal faulting
    and negative normal tractions correspond to compression. (b) Final
    slip as a function of depth in TPV13-2D for the triangular mesh
    with a resolution of 100 m on the fault.}
  \label{fig:tpv13-2d:stress:slip}
\end{figure}

Rupture initiates due to a low static coefficient of friction in the
nucleation region. Figure~\ref{fig:tpv13-2d:stress:slip}(a)
illustrates the depth dependence of the stress field in terms of the
fault tractions and Table~\ref{tab:tpv13:parameters} summarizes the
benchmark parameters.  \citet{Harris:etal:SRL:2011} provides a more
complete description with all of the details available from
\url{http://scecdata.usc.edu/cvws/cgi-bin/cvws.cgi}.  A challenging
feature of this, and many other benchmarks in the SCEC Spontaneous
Rupture Code Verification Exercise, is the use of parameters with
spatial variations that are not continuous. This includes the
variation in the static coefficient of friction for the nucleation
region and the transition to zero deviatoric stresses near the bottom
of the fault. We impose the geometry of these discontinuities in the
construction of the finite-element mesh and use the spatial average of
the parameters where they are discontinuous. This decreases the
sensitivity of the numerical solution to the discretization size. This
SCEC benchmark also includes fluid pressures. Because PyLith does not
include fluid pressure, we instead formulate the simulation parameters
in terms of effective stresses.

\begin{figure*}[h]
  \noindent\includegraphics{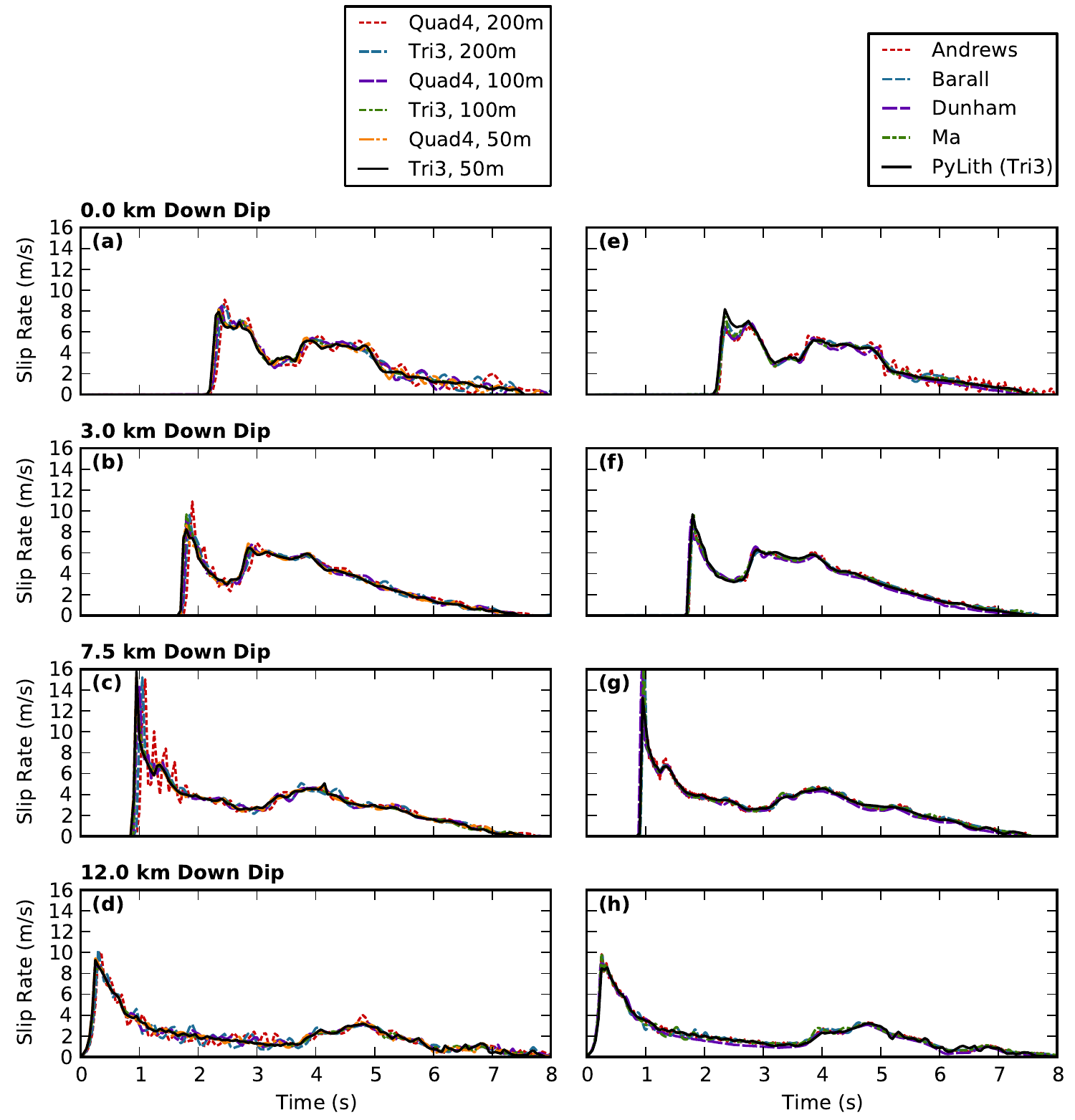}
  \caption{Slip rate time histories for SCEC spontaneous rupture benchmark
    TPV13-2D. Locations correspond to the red dots along the
    center-line of the fault shown in
    Figure~\ref{fig:tpv13:geometry}. Panels (a)--(d) show convergence
    of the solution for quadrilateral and triangular cells as a
    function of discretization size, and panels (e)--(h) demonstrate
    of code verification via excellent agreement among PyLith and four
    other dynamic rupture modeling codes
    \citep{Harris:etal:SRL:2011}.}
  \label{fig:tpv13-2d:slip:rate}
\end{figure*}

The TPV13-2D simulations require a small fraction of the computational
resources needed for the TPV13 3-D simulations and run quickly on a
laptop or desktop computer. The 50 m resolution cases took 62 s
(triangular cells) and 120 s (quadrilateral cells) using 8 processes
on a dual quad core desktop computer with Intel Xeon E5630 processors.
Figure~\ref{fig:tpv13-2d:stress:slip}(b) displays the final slip
distribution in the TPV13-2D simulation with triangular cells at a
resolution of 100 m. The large dynamic stress drop and supershear
rupture generate 20 m of slip at a depth of about 7
km. Figure~\ref{fig:tpv13-2d:slip:rate}(a)--(d) demonstrates the
convergence of the solution as the discretization size decreases as
evident in the normal faulting component of fault slip rate time
histories. For a resolution of 200 m on the fault, the solution
contains some high-frequency oscillation due to insufficient
resolution of the cohesive zone \citep{Rice:1993}. The finer meshes
provide sufficient resolution of the cohesive zone so there is very
little high-frequency oscillation in the slip rate time histories. The
triangular cells generate less oscillation compared with quadrilateral
cells.

In this benchmark without an analytical solution, as in all of the
exercises in the SCEC spontaneous rupture benchmark suite, we rely on
comparison with other dynamic spontaneous rupture modeling codes to
verify the numerical implementation in
PyLith. Figure~\ref{fig:tpv13-2d:slip:rate}(e)--(h) compares the slip
rate time histories from PyLith with four other codes (see
\citet{Harris:etal:SRL:2011}, \citet{Andrews:etal:2007},
\citet{Barall:2009}, \citet{Ma:2009}, and \citet{Dunham:etal:2011} for
a discussion of these other finite-element and finite-difference
codes).  The slip rate time histories agree very well, although some
codes yield more oscillation than others. We attribute this to
variations in the amount of numerical damping used in the various
codes.

\begin{figure*}[h]
  \noindent\includegraphics{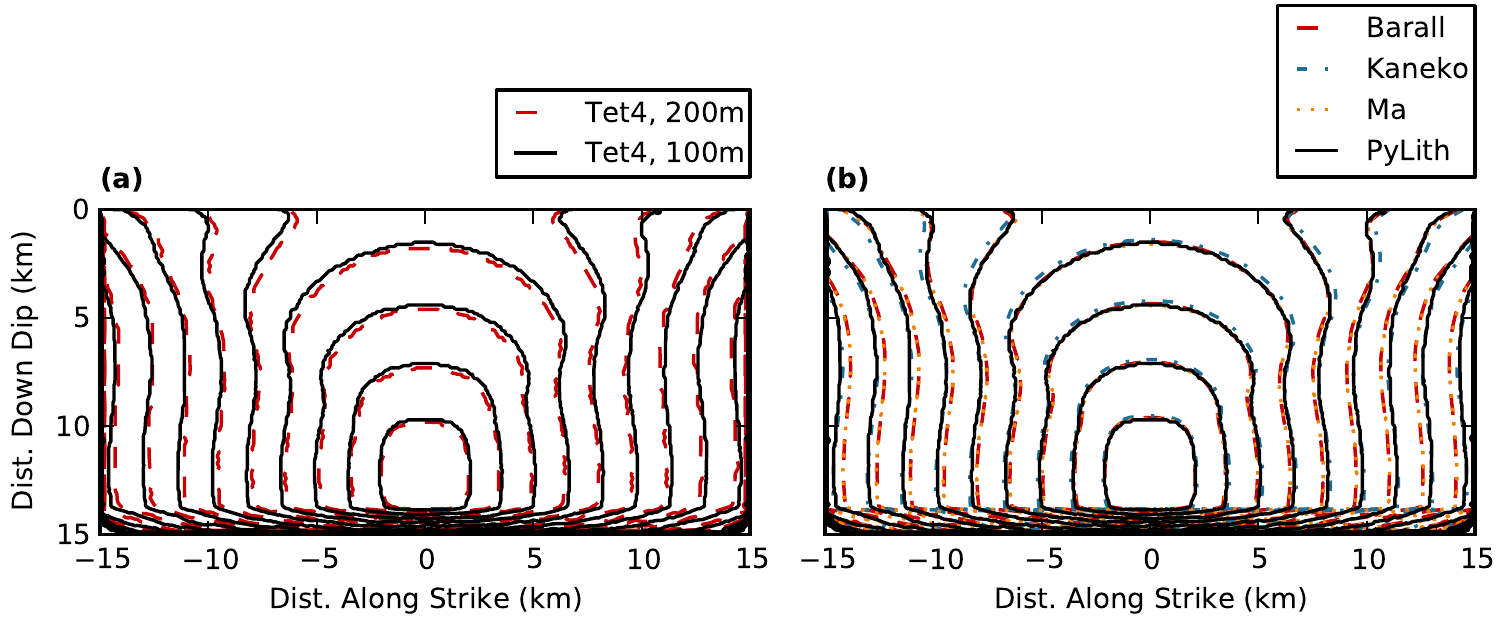}
  \caption{Rupture time contours (0.5 s interval) for SCEC spontaneous
    rupture benchmark TPV13. (a) Effect of discretization size and (b)
    demonstration of code verification via excellent agreement among
    PyLith and three other dynamic rupture modeling codes
    \citep{Harris:etal:SRL:2011}. The contours for PyLith and Kaneko
    (spectral element code) are nearly identical.}
  \label{fig:tpv13:rupture:time}
\end{figure*}

The 3-D version of the TPV13 benchmark yields similar results but
requires greater computational resources. The simulations with a
discretization size of 100 m took 2.5 hours using 64 processes (8
compute nodes with 8 processes per dual quad core compute node) on a
cluster with Intel Xeon E5620 processors.
Figure~\ref{fig:tpv13:rupture:time}(a) shows the same trends in
rupture speed with discretization size that we observed in the 2-D
version. In both cases models with insufficient resolution to resolve
the cohesive zone propagate slightly slower than models with
sufficient resolution. In this case the differences between the
rupture times for the 200 m and 100 m resolution tetrahedral meshes
are less than 0.1 seconds over the entire fault surface. Comparing the
rupture times among the modeling codes in
Figure~\ref{fig:tpv13:rupture:time}, we find that the four codes fall
into two groups. In the mode-III (along-strike) direction, PyLith and
the spectral element code by \citet{Kaneko:etal:2008} are essentially
identical while the finite-element codes by \citet{Barall:2009} and
\citet{Ma:Andrews:2010} are also essentially identical. In the mode-II
(up-dip) direction all four codes agree very closely. As in the 2-D
version, we attribute the differences among the codes not to the
numerical implementation but the treatment of discontinuities in the
spatial variation of the parameters. This explains why the
higher-order spectral element code by \citet{Kaneko:etal:2008} agrees
so closely with PyLith, a lower-order finite-element code.

\begin{figure*}[h]
  \noindent\includegraphics{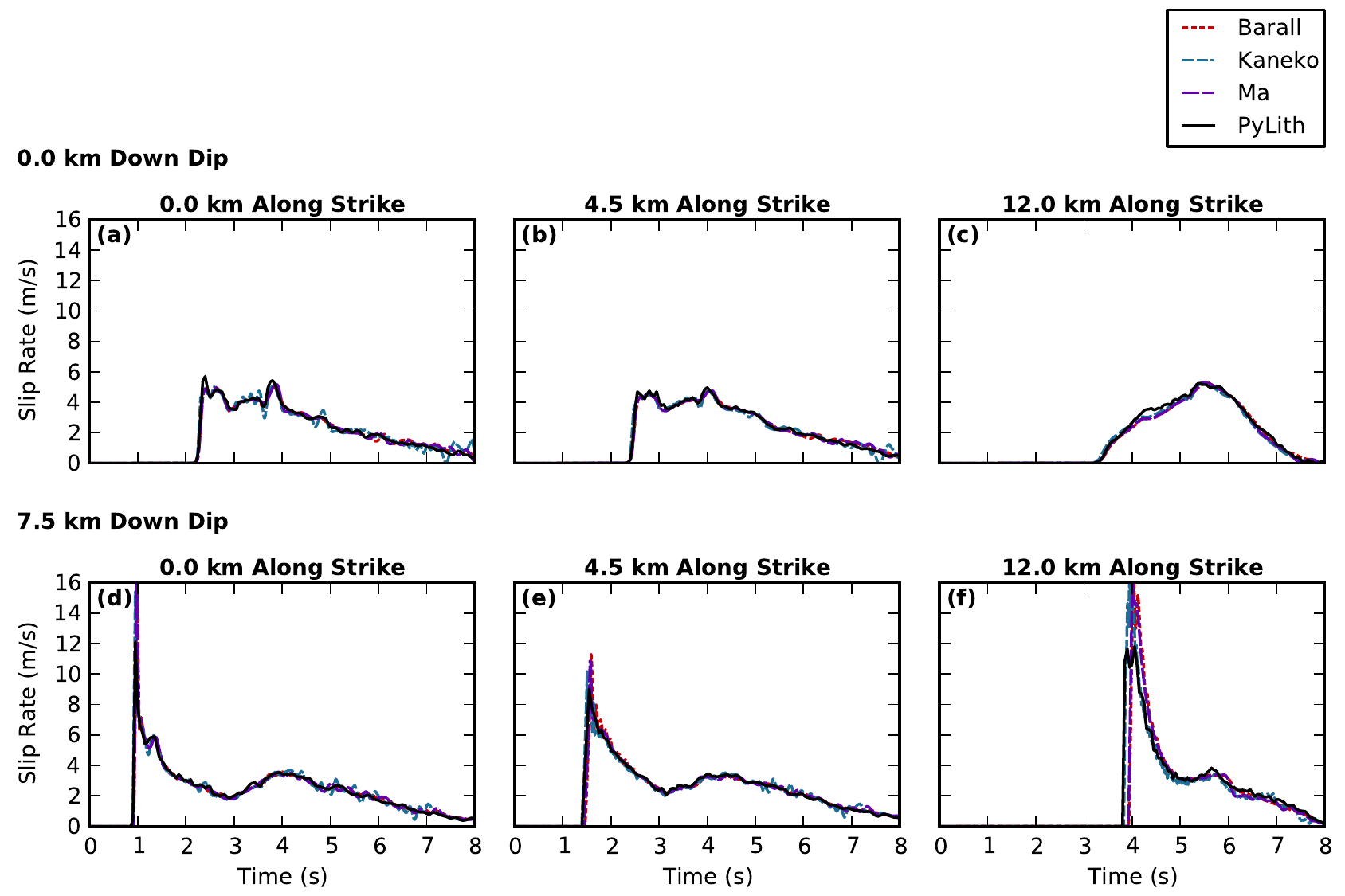}
  \caption{Comparison of normal faulting component of slip rate at six
    locations on the fault surface for SCEC spontaneous rupture benchmark
    TPV13. (a)--(c) are at a depth of 0 km and (d)--(f) are at a depth
    of 7.5 km. The slip rate time histories for all four dynamic
    rupture modeling codes agree very well. At 12 km along strike and
    7.5 km down dip, there is a small discrepancy between two groups
    of codes (PyLith and Kaneko versus Barall and Ma) that we
    attribute to how the modelers handled the discontinuity in the
    initial stress field and parameters.}
  \label{fig:tpv13:slip:rate}
\end{figure*}

\begin{figure*}[h]
  \noindent\includegraphics{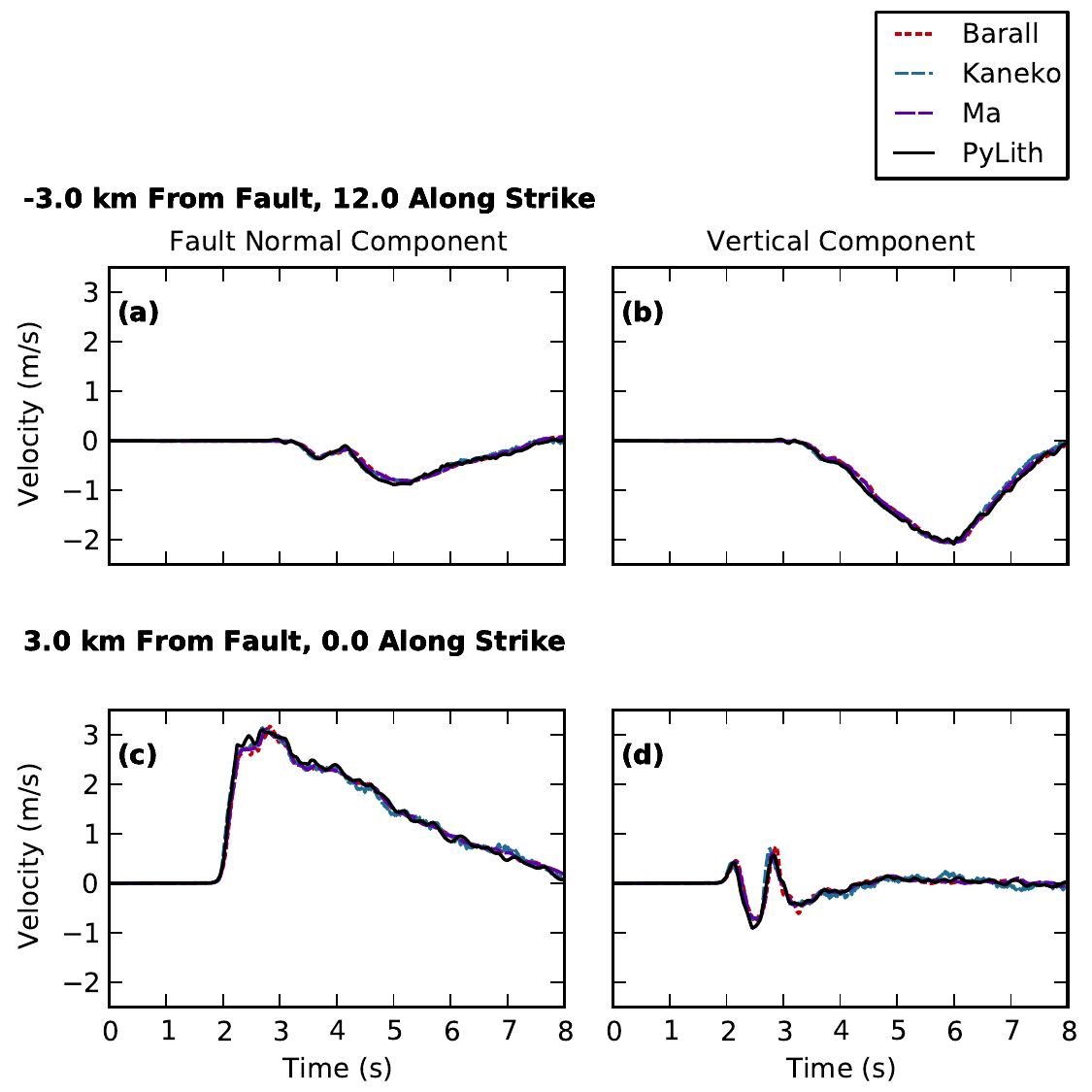}
  \caption{Comparison of fault normal and vertical components of
    velocity time histories at two sites on the ground surface for
    SCEC spontaneous rupture benchmark TPV13. Panels (a)--(b) are
    associated with a site that is on the hanging wall 3 km from the
    fault trace and 12 km along strike, and panels (c)--(d) are
    associated with a site that is on the footwall 3 km from the fault
    trace along the fault center-line. As expected based on the close
    agreement in the rupture time contours and fault slip rates, the
    velocity time histories from the difference dynamic rupture
    modeling codes agree very closely.}
  \label{fig:tpv13:velocity}
\end{figure*}

The slip rate and velocity time histories displayed in
Figures~\ref{fig:tpv13:slip:rate} and~\ref{fig:tpv13:velocity} are
consistent with the trends observed in the comparison of rupture
times. Furthermore, the codes all produce consistent results
throughout the entire time histories. The small differences in rupture
time in the mode-III (along-strike) direction between the two groups of
codes is evident in the slip rate time histories at a depth of 7.5 km
and 12 km along strike
(Figure~\ref{fig:tpv13:slip:rate}(f)). Nevertheless, this simply
produces a small time shift in the time history.

From the 2-D and 3-D versions of the SCEC spontaneous rupture
benchmark TPV13, we conclude that PyLith performs similarly
to other finite-element and finite-difference dynamic spontaneous
rupture modeling codes. In particular it is well-suited to problems
with complex geometry, as we are able to vary the discretization size
while simulating a dipping normal fault. The code accurately captures
supershear rupture and properly implements a Drucker-Prager
elastoplastic bulk rheology and slip-weakening friction.

\section{Conclusions}
\label{sec:conclusions}

PyLith provides a flexible numerical implementation of fault slip
using a domain decomposition approach. We have evaluated the
efficiency of several preconditioners for use of this fault
implementation in quasi-static simulations. We find that algebraic
multigrid preconditioners for elasticity combined with a custom
preconditioner for the fault block associated with the Lagrange
multipliers accelerates the convergence of the Krylov solver with the
fewest number of iterations and the least sensitivity to problem
size. Benchmark tests demonstrate the accuracy of our fault slip
implementation in PyLith with excellent agreement to (1) an analytical
solution for viscoelastic relaxation and strike-slip faulting over
multiple earthquake cycles and (2) other codes for supershear dynamic
spontaneous rupture on a dipping normal fault embedded in an
elastoplastic domain. Consequently, we believe this methodology
provides a promising avenue for modeling the earthquake cycle through
coupling of quasi-static simulations of the interseismic and
postseismic deformation and dynamic rupture simulations of earthquake
rupture propagation.

\begin{notation}
  $\bm{A}$ & matrix associated with Jacobian operator for the entire system of equations.\\
  $\bm{C}$ & fourth order tensor of elastic constants.\\
  $\bm{d}$ & fault slip vector.\\
  $\bm{f}$ & body force vector.\\
  $\bm{l}$ & Lagrange multiplier vector corresponding to the fault traction vector.\\
  $\bm{L}$ & matrix associated with Jacobian operator for constraint equation.\\
  $\bm{K}$ & matrix associated with Jacobian operator for
  elasticity equation.\\
  $\bm{N}_m$ & matrix for $m$ basis functions.\\
  $\bm{n}$ & normal vector.\\
  $\bm{P}$ & preconditioning matrix.\\
  $\bm{P}_\mathit{elastic}$ & preconditioning matrix associated with elasticity.\\
  $\bm{P}_\mathit{fault}$ & preconditioning matrix associated with fault slip constraints (Lagrange multipliers).\\
  $S_f$ & fault surface.\\
  $S_T$ & surface with Neumann boundary conditions.\\
  $S_u$ & surface with Dirichlet boundary conditions.\\
  $t$ & time.\\
  $\bm{T}$ & Traction vector.\\
  $T_c$ & scalar shear traction associated with cohesion.\\
  $T_f$ & scalar shear traction associated with friction.\\
  $T_n$ & scalar normal traction.\\
  $\bm{u}$ & displacement vector.\\
  $V$ & spatial domain of model.\\
  $V_p$ & dilatational wave speed. \\
  $V_s$ & shear wave speed.\\
  $\Delta t$ & time step.\\
  $\eta^{*}$ & nondimensional viscosity used for numerical damping.\\
  $\pmb{\phi}$ & weighting function.\\
  $\mu_f$ & coefficient of friction.\\
  $\rho$ & mass density.\\
  $\bm{\sigma}$ & Cauchy stress tensor.
\end{notation}

\begin{acknowledgments}
  We thank Sylvain Barbot, Ruth Harris, and Fred Pollitz for their
  careful reviews of the manuscript. Development of PyLith has been
  supported by the Earthquake Hazards Program of the U.S. Geological
  Survey, the Computational Infrastructure for Geodynamics (NSF grant
  EAR-0949446), GNS Science, and the Southern California Earthquake
  Center. SCEC is funded by NSF Cooperative Agreement EAR-0529922 and
  USGS Cooperative Agreement 07HQAG0008. PyLith development has also
  been supported by NSF grants EAR/ITR-0313238 and EAR-0745391. This
  is SCEC contribution number 1665. Several of the figures were
  produced using Matplotlib \citep{matplotlib} and PGF/TikZ (available
  from \url{http://sourceforge.net/projects/pgf/}). Computing
  resources for the parallel scalability benchmarks were provided by
  the Texas Advanced Computing Center (TACC) at The University of
  Texas at Austin (\url{http://www.tacc.utexas.edu}).
\end{acknowledgments}


\pagebreak

\pagebreak

\clearpage

\clearpage

\begin{table}
\scriptsize 
  \caption{Example Preconditioners for the Saddle Point Problem in
    Equation~(\ref{eqn:saddle:point})\tablenotemark{a}}
  \label{tab:preconditioner:options}
\centering
\begin{tabular}{ll}
  AMG with additive relaxation                         & AMG with multiplicative relaxation \\
  $\begin{pmatrix}\bm{K} & \bm{0} \\ \bm{0} & \bm{I}\end{pmatrix}$ & $\begin{pmatrix}\bm{K} & \bm{L}^T \\ \bm{0} & \bm{I}\end{pmatrix}$\\
  \texttt{[pylithapp.problem.formulation]}             & \texttt{[pylithapp.problem.formulation]} \\
  \texttt{split\_fields = True}                        & \texttt{split\_fields = True} \\
  \texttt{matrix\_type = aij}                          & \texttt{matrix\_type = aij} \\
  \texttt{[pylithapp.petsc]}                           & \texttt{[pylithapp.petsc]} \\
  \texttt{fs\_pc\_type = fieldsplit}                   & \texttt{fs\_pc\_type = fieldsplit} \\
  \texttt{fs\_pc\_field\_split\_real\_diagonal = true} & \texttt{fs\_pc\_field\_split\_real\_diagonal = true} \\
  \texttt{fs\_pc\_field\_split\_type = additive}       & \texttt{fs\_pc\_field\_split\_type multiplicative} \\
  \texttt{fs\_fieldsplit\_0\_pc\_type = ml}            & \texttt{fs\_fieldsplit\_0\_pc\_type = ml} \\
  \texttt{fs\_fieldsplit\_0\_ksp\_type = preonly}      & \texttt{fs\_fieldsplit\_0\_ksp\_type = preonly} \\
  \texttt{fs\_fieldsplit\_1\_pc\_type = jacobi}        & \texttt{fs\_fieldsplit\_1\_pc\_type = jacobi} \\
  \texttt{fs\_fieldsplit\_1\_ksp\_type = gmres}        & \texttt{fs\_fieldsplit\_1\_ksp\_type = gmres} \\
  \smallskip \\
  Schur complement, upper factorization                       & Schur complement, full factorization \\
  $\begin{pmatrix}\bm{K} & \bm{L}^T \\ \bm{0} & \bm{S}\end{pmatrix}$ & $\begin{pmatrix}\bm{I} & \bm{0} \\ \bm{B}^T \bm{A}^{-1} & \bm{I}\end{pmatrix}\begin{pmatrix}\bm{A} & \bm{0} \\ \bm{0} & \bm{S}\end{pmatrix}\begin{pmatrix}\bm{I} & \bm{A}^{-1} \bm{B} \\ \bm{0} & \bm{I}\end{pmatrix}$ \\
  \texttt{[pylithapp.problem.formulation]}                    & \texttt{[pylithapp.problem.formulation]} \\
  \texttt{split\_fields = True}                               & \texttt{split\_fields = True} \\
  \texttt{matrix\_type = aij}                                 & \texttt{matrix\_type = aij} \\
  \texttt{[pylithapp.petsc]}                                  & \texttt{[pylithapp.petsc]} \\
  \texttt{pc\_type = fieldsplit}                              & \texttt{pc\_type = fieldsplit} \\
  \texttt{pc\_field\_split\_type = schur}                     & \texttt{pc\_field\_split\_type = schur} \\
  \texttt{fs\_pc\_field\_split\_real\_diagonal = true}        & \texttt{fs\_pc\_field\_split\_real\_diagonal = true} \\
  \texttt{pc\_fieldsplit\_schur\_factorization\_type = upper} & \texttt{pc\_fieldsplit\_schur\_factorization\_type = full} \\
  \texttt{pc\_fieldsplit\_schur\_precondition = user}         & \texttt{pc\_fieldsplit\_schur\_precondition = user} \\
  \texttt{fieldsplit\_0\_pc\_type = ml}                       & \texttt{fieldsplit\_0\_pc\_type = ml} \\
  \texttt{fieldsplit\_0\_ksp\_type = preonly}                 & \texttt{fieldsplit\_0\_ksp\_type = preonly} \\
  \texttt{fieldsplit\_1\_pc\_type = jacobi}                   & \texttt{fieldsplit\_1\_pc\_type = jacobi} \\
  \texttt{fieldsplit\_1\_ksp\_type = gmres}                   & \texttt{fieldsplit\_1\_ksp\_type = gmres} \\
\end{tabular}
\tablenotetext{a}{Four examples of preconditioners often used to
  accelerate convergence in saddle point problems. Below the
  mathematical expression for the preconditioner, we show the PyLith
  parameters used to construct the preconditioner. }
\end{table}

\clearpage
\begin{table}
\caption{Performance Benchmark Parameters\tablenotemark{a}}
\label{tab:solvertest:parameters}
\centering
\begin{tabular}{llc}
  \hline
  \multicolumn{2}{l}{Parameter} & Value \\
  \hline
  \multicolumn{2}{l}{Domain} & \\
    & Length & 72 km \\
    & Width & 72 km \\
    & Height & 36 km \\
    & Angle between faults & 60 $\deg$ \\
  \multicolumn{2}{l}{Elastic properties} & \\
    & Vp & 5.774 km/s \\
    & Vs & 3.333 km/s \\
    & Density ($\rho$) & 2700. kg/m$^3$ \\
  \multicolumn{2}{l}{Middle fault} & \\
    & Length & 39.19 km \\
    & Width & 12 km \\
    & Slip & 1.0 m RL \\
  \multicolumn{2}{l}{End faults} & \\
    & Length & 43.74 km \\
    & Width & 12 km \\
    & Slip & 0.5 m LL \\
  \hline
\end{tabular}
\tablenotetext{a}{Simulation parameters for the performance benchmark
  with three faults embedded in a volume domain as shown in
  Figure~\ref{fig:solvertest:geometry}. We prescribe right-lateral
  (RL) slip on the middle fault and left-lateral (LL) slip on the end faults.}
\end{table}

\begin{table}
\caption{Preconditioner Performance\tablenotemark{a}}
\label{tab:solvertest:preconditioner:iterates}
\centering
\begin{tabular}{lcrrr}
  \hline
  Preconditioner & Cell & \multicolumn{3}{c}{Problem Size} \\
     &      & S1 & S2 & S4 \\
  \hline
  ASM
    & Tet4 & 184 & 217 & 270 \\
    & Hex8 & 143 & 179 & 221 \\
  Schur (full)
    & Tet4 & 82 & 84 & 109 \\
    & Hex8 & 54 & 60 & 61 \\
  Schur (upper)
    & Tet4 & 79 & 78 & 87 \\
    & Hex8 & 53 & 59 & 57 \\
  FieldSplit (add)
    & Tet4 & 241 & 587 & 585 \\
    & Hex8 & 159 & 193 & 192 \\
  FieldSplit (mult)
    & Tet4 & 284 & 324 & 383 \\
    & Hex8 & 165 & 177 & 194 \\
  FieldSplit (mult,custom)
    & Tet4 & 42 & 48 & 51 \\
    & Hex8 & 35 & 39 & 43 \\
  \hline
\end{tabular}
\tablenotetext{a}{Number of iterations for additive Schwarz (ASM),
  Schur complement (Schur), and field split (additive, multiplicative,
  and multiplicative with custom fault block preconditioner),
  preconditioners for tetrahedral and hexahedral discretizations and
  three problem sizes (S1 with $1.8\times 10^5$ DOF, S2 with
  $3.5\times 10^5$ DOF, and S3 with $6.9\times 10^5$ DOF). The Schur
  complement preconditioners and the field split preconditioner with
  multiplicative factorization and the custom fault block
  preconditioner yield the best performance with only a 
  fraction of the iterates as the other preconditioners and a small
  increase with problem size. Furthermore, the field
  split preconditioner with multiplicative factorization and the custom
  fault block preconditioner provides the shortest runtime.}
\end{table}

\begin{table}
\scriptsize 
\caption{Performance Benchmark Memory System Evaluation\tablenotemark{a}}
\label{tab:solvertest:memory:events}
\centering
\begin{tabular}{lrrr}
  \hline
  Event & \# Cores & Load Imbalance & MFlops/s \\
  \hline
VecMDot &    1 & 1.0 &   2007 \\
     &    2 & 1.1 &   3809 \\
     &    4 & 1.1 &   5431 \\
     &    6 & 1.1 &   5967 \\
     &   12 & 1.2 &   5714 \\
     &   24 & 1.2 &  11784 \\
     &   48 & 1.2 &  20958 \\
     &   96 & 1.3 &  17976 \\
  \hline
VecAXPY &    1 & 1.0 &   1629 \\
     &    2 & 1.1 &   3694 \\
     &    4 & 1.1 &   5969 \\
     &    6 & 1.1 &   6028 \\
     &   12 & 1.2 &   5055 \\
     &   24 & 1.2 &  10071 \\
     &   48 & 1.2 &  18761 \\
     &   96 & 1.3 &  33676 \\
  \hline
VecMAXPY &    1 & 1.0 &   1819 \\
     &    2 & 1.1 &   3415 \\
     &    4 & 1.1 &   5200 \\
     &    6 & 1.1 &   5860 \\
     &   12 & 1.2 &   6051 \\
     &   24 & 1.2 &  12063 \\
     &   48 & 1.2 &  23072 \\
     &   96 & 1.3 &  28461 \\
  \hline
\end{tabular}
\tablenotetext{a}{Examination of memory system performance using three
  PETSc vector operations for simulations with the hexahedral
  meshes. The performance for the tetrahedral meshes is very similar. For ideal scaling the number of floating point operations
  per second should scale linearly with the number of processes. \texttt{VecMDot}
  corresponds to the operation for vector reductions, \texttt{VecAXPY}
  corresponds to vector scaling and addition, and \texttt{VecMAXPY}
  corresponds to multiple vector scaling and addition.}
\end{table}

\begin{table}
\scriptsize
\caption{Performance Benchmark Solver Evaluation\tablenotemark{a}}
\label{tab:solvertest:solver:events}
\centering
\begin{tabular}{lrrr}
  \hline
  Event & \# Calls & Time (s) & MFlops/s \\
  \hline
\multicolumn{4}{c}{{\it p = 12}} \\
  MatMult & 180 &      2.7 &     6113 \\
  PCSetUp &   1 &      5.7 &      232 \\
  PCApply &  57 &      5.5 &     3690 \\
  KSPSolve &   1 &     15.1 &     3013 \\
\hline
\multicolumn{4}{c}{{\it p = 24}} \\
  MatMult & 207 &      3.1 &    12293 \\
  PCSetUp &   1 &      5.2 &      526 \\
  PCApply &  66 &      6.6 &     7285 \\
  KSPSolve &   1 &     16.4 &     6666 \\
\hline
\multicolumn{4}{c}{{\it p = 48}} \\
  MatMult & 222 &      4.0 &    21136 \\
  PCSetUp &   1 &     10.1 &      628 \\
  PCApply &  71 &      9.4 &    12032 \\
  KSPSolve &   1 &     25.1 &    10129 \\
\hline
\multicolumn{4}{c}{{\it p = 96}} \\
  MatMult & 234 &      4.0 &    42130 \\
  PCSetUp &   1 &     11.8 &     1943 \\
  PCApply &  75 &     11.6 &    20422 \\
  KSPSolve &   1 &     30.5 &    17674 \\
\hline
\end{tabular}
\tablenotetext{a}{Examination of solver performance using three of the
  main events comprising the linear solve for simulations with the
  hexahedral meshes and 12, 24, 48, and 96 processes. The performance
  for the tetrahedral meshes is nearly the same. For ideal scaling
  the time for each event should be constant as the number of
  processes increases. The \texttt{KSPSolve} event encompasses the
  entire linear solve. \texttt{MatMult} corresponds to matrix-vector
  multiplications. \texttt{PCSetUp} and \texttt{PCApply} correspond to
  the setup and application of the AMG preconditioner.}
\end{table}

\begin{table}
\caption{Savage and Prescott Benchmark Parameters\tablenotemark{a}}
\label{tab:Savage:Prescott:parameters}
\centering
\begin{tabular}{llc}
  \hline
  \multicolumn{2}{l}{Parameter} & Value \\
  \hline
  \multicolumn{2}{l}{Domain} & \\
    & Length & 2000 km \\
    & Width & 1000 km \\
    & Height & 400 km \\
    & Fault dip angle & 90 $\deg$ \\
  \multicolumn{2}{l}{Physical properties} & \\
    & Shear modulus & 30 GPa \\
    & Viscosity & $2.37\times10^{19}$ Pa-s \\
    & Elastic thickness & 40 km \\
  \multicolumn{2}{l}{Fault slip} & \\
    & Locking depth & 20 km \\
    & Earthquake recurrence & 200 yr \\
    & Earthquake slip & 4 m \\
    & Creep rate & 2 cm/yr \\
  \hline
\end{tabular}
\tablenotetext{a}{Simulation parameters for the Savage and Prescott benchmark
  with multiple earthquake cycles on a vertical strike-slip fault
  embedded in an elastic layer over a viscoelastic half-space.}
\end{table}

\begin{table}
\caption{SCEC Benchmark TPV13 Parameters\tablenotemark{a}}
\label{tab:tpv13:parameters}
\centering
\begin{tabular}{llc}
  \hline
  \multicolumn{2}{l}{Parameter} & Value \\
  \hline
  \multicolumn{2}{l}{Domain} & \\
    & Length & 64 km \\
    & Width & 48 km \\
    & Height & 36 km \\
    & Fault dip angle & 60 $\deg$ \\
  \multicolumn{2}{l}{Elastic properties} & \\
    & Vp & 5.716 km/s \\
    & Vs & 3.300 km/s \\
    & Density ($\rho$) & 2700. kg/m$^3$ \\
    & Nondimensional viscosity ($\eta^*$) & 0.4 \\ 
  \hline
\end{tabular}
\tablenotetext{a}{Basic simulation parameters for the SCEC
  dynamic spontaneous rupture benchmark TPV13. A complete list of
  the parameters can be found in \citet{Harris:etal:SRL:2011}.}
\end{table}

\clearpage

\end{article}

\end{document}